\documentclass[10pt]{article}
\usepackage{fullpage}
\usepackage{amsmath}
\usepackage{amssymb}
\usepackage{amsfonts}
\usepackage[dvips]{epsfig}
\usepackage{color}

\def\##1{\underline{#1}}
\def\=#1{\underline{\underline{#1}}}

\def\+#1{\underline{\bf #1}}
\def\*#1{\underline{\underline{\bf #1}}}

\def\r#1{(\ref{#1})}
\def\l#1{\label{#1}}
\def\c#1{\cite{#1}}

\def\le{\left(}
\def\ri{\right)}
\def\les{\left[}
\def\ris{\right]}
\def\lec{\left\{}
\def\ric{\right\}}

\def\.{\mbox{ \tiny{$^\bullet$} }}

\def\eps{\varepsilon}

\def\epso{\eps_{\scriptscriptstyle 0}}
\def\lambdao{\lambda_{\scriptscriptstyle 0}}

\def\co{c_{\scriptscriptstyle 0}}

\def\ux{\hat{\#u}_x}
\def\uy{\hat{\#u}_y}
\def\uz{\hat{\#u}_z}

\begin{document}

\begin{center}

\LARGE{ {\bf Application of Bruggeman and Maxwell Garnett homogenization formalisms to random composite materials containing dimers
}}
\end{center}
\begin{center}
\vspace{10mm} \large

 Tom G. Mackay\footnote{E--mail: T.Mackay@ed.ac.uk.}\\
{\em School of Mathematics and
   Maxwell Institute for Mathematical Sciences\\
University of Edinburgh, Edinburgh EH9 3FD, UK}\\
and\\
 {\em NanoMM~---~Nanoengineered Metamaterials Group\\ Department of Engineering Science and Mechanics\\
Pennsylvania State University, University Park, PA 16802--6812,
USA}\\
 \vspace{3mm}
 Akhlesh  Lakhtakia\footnote{E--mail: akhlesh@psu.edu}\\
 {\em NanoMM~---~Nanoengineered Metamaterials Group\\ Department of Engineering Science and Mechanics\\
Pennsylvania State University, University Park, PA 16802--6812, USA}

\normalsize

\end{center}

\begin{center}
\vspace{15mm} {\bf Abstract}

\end{center}
The homogenization of a  composite material comprising  three isotropic dielectric materials was investigated. The component materials were
randomly distributed as spherical particles, with the particles of two of the component materials being coupled to form dimers.
The Bruggeman and Maxwell Garnett formalisms were developed to estimate the  permittivity dyadic of the  homogenized composite material (HCM), under the quasi-electrostatic approximation. Both randomly oriented and identically oriented dimers were accommodated; in the former case the HCM is  isotropic, whereas in the latter case the HCM is uniaxial. Representative numerical results for composite materials containing dielectric--dielectric dimers demonstrate close agreement between the estimates   delivered by the Bruggeman and Maxwell Garnett formalisms. For
composite materials containing metal--dielectric dimers with moderate degrees of dissipation, the estimates of the two formalisms are in broad agreement, provided that the dimer volume fractions  are relatively low. In general, the effects of intradimer coupling on the estimates of the HCM permittivity are relatively modest but not insignificant, these effects being exacerbated by anisotropy when all dimers are identically oriented.

\vspace{4mm}


\vspace{4mm}

\section{Introduction}

Composite materials containing random dispersals of particles or inclusions
can exhibit remarkable characteristics if two (or more) of their component materials
are coupled together to form dimers (or trimers, etc.). Indeed,  nanoengineered composite
materials containing dimeric particles are playing an increasingly prominent role in the development of new technologies \c{Nanoengineering}. For example, plasmonic interactions in metallic dimers can result
in enhanced Raman scattering or molecular fluorescence \c{Lombardi,Park,Nordlander}, which may lead to  highly sensitive optical sensors. Dielectric dimers are also of considerable interest,  in the context of Mott insulators \c{Mott_insulator_PRL, Mott_insulator_PRB}, liquid crystals \c{Dimer_Liquid_Crystals}, and interstellar molecular hydrogen formation \c{Hydrogen_dimer},  for examples.

This study concerns the estimation of the effective constitutive parameters of particulate composite materials,
wherein two of the component materials are jointly present as dimers. Previously this topic has been investigated
 using Mie scattering theory \c{Zhang}, a Brownian-motion formulation \c{Torquato},
 numerical methods such as the finite element method \c{Lombardi}, and quantum-mechanical methods based on density function theory \c{Nordlander,Hydrogen_dimer}. In contrast,
the theoretical approaches taken herein are simpler, being based
on the well-established homogenization formalisms named after Bruggeman and Maxwell Garnett \c{L96,ML_PiO}.
The Bruggeman formalism is a self-consistent one in which all component materials are treated in an
identical manner. A rigorous basis for the Bruggeman formalism arises from the strong-permittivity-fluctuation theory \c{TK81}. The nature of the Maxwell Garnett formalism is essentially perturbative---which is only
appropriate for dilute dispersals
of particles in a host  material \c{Faxen,Lakh1992}. The theoretical basis for the Maxwell Garnett formalism is bolstered by its close association with the Hashin--Shtrikman bounds \c{H-S_bounds}.

The plan of this paper is as follows. Relevant details of the homogenization formalisms are provided in \S\ref{Homog_prelim}. This is followed by the development of appropriate expressions for the electric dipole moments and polarizability densities in \S\ref{dipole_sec}. The homogenization formalisms themselves are set up for composite materials
containing
identically oriented dimers in \S\ref{Sec_UOD} and randomly oriented dimers in \S\ref{Sec_ROD}. Representative numerical results  are presented
in \S\ref{Num_sec} for both dielectric--dielectric dimers  and metal--dielectric dimers. Lastly, a brief discussion of the theory and numerical results is provided in \S\ref{Dis_sec}. As regards notation: the permittivity of free space is written as $\epso$;  $\co$ denotes the speed of light in free space; $\omega$ is the angular frequency; and $i = \sqrt{-1}$. Vectors are underlined (with the $\hat{}$ symbol denoting unit vectors), whereas dyadics \cite{Chen,EAB} are double underlined.

\section{Homogenization preliminaries} \l{Homog_prelim}

The homogenization of three  isotropic component materials, labeled as `a', `b',  and `c',
of a particulate composite material is investigated in the following sections.
The three component materials are characterized by the  permittivity scalars $\eps_{a}$, $\eps_{b}$, and $\eps_{c}$.

Component materials `a' and `b' are dispersed as dimers of electrically small spheres. For simplicity, the average radiuses of the
spheres belonging to these two component materials are taken to be the same, namely $s$.
Each sphere of component material `a'  is chemically linked to exactly one sphere of component material `b',
thus forming a dimer.
The distance between the centers of the two  spheres making up a dimer is $d \geq 2 s $.
In each dimer,  the location of the center of the `b' sphere  relative to the center of the `a' sphere
is given by $d \#{\hat{d}}$,  where the unit vector
\begin{equation}
 \#{\hat{d}} =    \left(\ux\cos \phi + \uy \sin \phi\right) \sin \theta  +\uz\cos\theta , \qquad \theta \in \les \, 0, \pi \ris, \quad \phi \in \les \, 0 , 2 \pi \ri.
\end{equation}
 A heterodimer is specified by $\eps_{a} \neq \eps_{b}$, whereas a homodimer is specified by $\eps_{a} = \eps_{b}$.
The  dimers are randomly distributed.

When applying the Bruggeman  formalism, component material `c' is regarded as a random dispersal of electrically small spheres,
which we take to have an average radius $s$. No particular topology is assigned to component material `c'.
in the Maxwell Garnett  formalism.
 The volume fractions of component materials `a', `b', and `c' are $f_a$, $f_b$, and $f_c$, respectively, with
 $f_a = f_b$ and $f_a + f_b + f_c = 1$.

Provided that the largest relevant wavelength is much longer than the linear dimensions of the spheres and the
dimers, the mixture of component materials `a', `b', and `c' may be
regarded as being effectively homogeneous. The constitutive parameters of the resulting homogenized composite material (HCM) may be estimated using a homogenization formalism. Two distinct cases are investigated theoretically:  identically oriented dimers are the subject of  \S\ref{Sec_UOD}, while randomly oriented dimers are treated in \S\ref{Sec_ROD}.

 All dimers are assumed to be oriented in the same direction in  \S\ref{Sec_UOD}; i.e., $ \#{\hat{d}}$ is fixed for all dimers.
Consequently, the  HCM is a uniaxial  dielectric material characterized by a  permittivity dyadic
of the form
 \begin{equation} \l{e_HCM}
 \=\eps_{\,HCM} = \eps_{HCM}^\perp \le \=I - \#{\hat{d}} \, \#{\hat{d}} \ri + \eps_{HCM}^\parallel \#{\hat{d}} \, \#{\hat{d}} \,,
 \end{equation}
 where $\=I$ is the identity dyadic.
 The estimate of $\=\eps_{\,HCM}$ (with components $\eps_{HCM}^{\perp}$ and $\eps_{HCM}^{\parallel}$)  yielded by the Bruggeman formalism is written as
 $\=\eps_{\,Br}$ (with components  $\eps^{\perp}_{Br}$ and $\eps^{\parallel}_{Br}$) and the estimate  yielded by the Maxwell Garnett formalism is written as  $\=\eps_{\,MG}$ (with components $\eps^{\perp}_{MG}$ and $\eps^{\parallel}_{MG}$).

 As the dimers are taken to have no preferred orientation in  \S\ref{Sec_ROD}, the corresponding HCM is an isotropic dielectric material characterized by the scalar permittivity $\eps_{HCM}$. The estimate of  $\eps_{HCM}$ yielded by the Bruggeman formalism is written as  $\eps_{Br}$, and the estimate  yielded by the Maxwell Garnett formalism is written as  $\eps_{MG}$.

\section{Electric dipole moments and polarizability densities} \l{dipole_sec}

A quasi-electrostatic viewpoint is adopted in both homogenization formalisms,
whereby each pair of  electrically small spheres comprising a
dimer is modeled as a pair of electric dipole moments  $\#p_{\,a}$ and $\#p_{\,b}$, separated by the distance $d$ in the direction of $\#{\hat{d}}$. In the case of the Bruggeman formalism, each electrically small sphere of  component `c'  is  modeled as a electric dipole moment  $\#p_{\,c}$. No  magnetic dipole moments analogous to $\#p_{\,a,b,c}$   are present in the quasi-electrostatic regime.

We need to consider two separate cases. In \S\ref{uniaxial_out_sec} the electrically small spheres are immersed
in a homogeneous uniaxial dielectric material, while in
\S\ref{isotropic_out_sec} the electrically small spheres are immersed
in a homogeneous isotropic dielectric material.

\subsection{Uniaxial dielectric immersion material} \l{uniaxial_out_sec}

Let us consider a single dimer  immersed in a homogeneous dielectric material characterized by the permittivity dyadic $\=\eps_{\,out}$.
 In this section, $\=\eps_{\,out}$  characterizes a uniaxial material whose distinguished axis is parallel to $\#{\hat{d}}$ ; i.e.,
 \begin{equation} \l{eps_out}
 \=\eps_{\,out} = \eps^\perp_{out} \le \=I - \#{\hat{d}} \, \#{\hat{d}} \ri + \eps^\parallel_{out}\, \#{\hat{d}} \, \#{\hat{d}} \,.
 \end{equation}
 The dimer is taken to be centered at the origin $\#r = \#0$.

 A  quasi-electrostatic  field, originating from a distant source, is  incident on this dimer.  The incident electric field phasor is denoted by $\#E_{\,inc} (\#r) $.
In response to  $\#E_{\,inc} (\#r)$, electric dipole moments $\#p_{\,a}$ and $\#p_{\,b}$ are
 induced at $\#r_{\,a} = - \le d/2 \ri \#{\hat{d}}$
and $\#r_{\,b} = \le d/2 \ri \#{\hat{d}}$, respectively.
These electric dipole moments are given by \c{Jackson}
\begin{equation} \l{p_a_dipole}
\#p_{\,\ell} = \frac{4}{3} \pi  s^3 \=\alpha^{}_{\, \ell/out} \.  \#E_{\,exc} (\#r_{\, \ell}), \qquad   \ell \in\lec a,b \ric,
\end{equation}
where
\begin{equation} \l{a_polarizability_def}
\=\alpha^{}_{\, \ell/out} = \le \eps_{ \ell } \=I - \=\eps_{\,out} \ri \. \les \=I + i \omega \=D_{\,out} \.
\le \eps_{\ell } \=I - \=\eps_{\,out} \ri
\ris^{-1}
\end{equation}
 is the
  polarizability density dyadic of an isolated  sphere of material `$\ell$', immersed in the material labeled `out', with
   $\=D_{\,out}$ being the corresponding depolarization dyadic \cite{EAB}.

The 3$\times$3 dyadics $\=D_{\,out}$ and $\=\alpha^{}_{\,\ell/out}$ possess the same symmetric form as $\=\eps_{\,out}$, i.e.,
\begin{equation} \l{D_alpha_form}
\left.
\begin{array}{l}
\=D^{}_{\,out} =  D^{\perp}_{out} \le \=I - \#{\hat{d}} \, \#{\hat{d}} \ri + D^{\parallel}_{out}\, \#{\hat{d}} \, \#{\hat{d}} \vspace{6pt}\\
\=\alpha^{}_{\, \ell/out} =  \alpha^{\perp}_{ \ell/out} \le \=I - \#{\hat{d}} \, \#{\hat{d}} \ri + \alpha^{\parallel}_{ \ell/out}\, \#{\hat{d}} \, \#{\hat{d}}
\end{array}
\right\}.
\end{equation}
The components of $\=D^{}_{\,out}$ are given as \c{Michel}
\begin{equation}
\left.
\begin{array}{l}
\displaystyle{D^{\perp}_{out} = \frac{\gamma}{i \omega \eps^\perp_{out}} \, L^{\perp} (\gamma)} \vspace{6pt} \\
\displaystyle{D^{\parallel}_{out} =\frac{1}{i \omega \eps^\perp_{out}} \, L^{\parallel} (\gamma)}
\end{array}
\right\},
\end{equation}
with
\begin{equation}
\left.
\begin{array}{l}
\displaystyle{ L^{\perp} (\gamma) = \frac{1}{2 \gamma} \les  \frac{1}{ \sqrt{1 - \gamma }} \tanh^{-1} \le \sqrt{1- \gamma } \ri - L^\parallel (\gamma) \ris }
\vspace{6pt} \\
\displaystyle{L^{\parallel} (\gamma) = \frac{1}{\gamma - 1} \les 1 -  \frac{1}{ \sqrt{1 - \gamma }} \tanh^{-1} \le \sqrt{1- \gamma } \ri \ris }
\end{array}
\right\}
\end{equation}
being dimensionless scalar functions of the dimensionless parameter $\gamma = \eps^\parallel_{out} / \eps^\perp_{out}$.
The components of $\=\alpha^{}_{\,\ell / out}$ are given as
\begin{equation}
\left.
\begin{array}{l}
\displaystyle{ \alpha^{\perp}_{ \ell/out} =  \frac{\eps^{\perp}_{out} \le \eps_{ \ell} - \eps^{\perp}_{out} \ri  }{\eps^{\perp}_{out} + \gamma L^\perp (\gamma) \le \eps_{ \ell} - \eps^{\perp}_{out} \ri }} \vspace{6pt} \\
\displaystyle{ \alpha^{\parallel}_{ \ell/out} =  \frac{\eps^{\perp}_{out} \le \eps_{ \ell} - \eps^{\parallel}_{out} \ri  }{\eps^{\perp}_{out} +  L^\parallel (\gamma) \le \eps_{ \ell} - \eps^{\parallel}_{out} \ri }}
\end{array}
\right\}.
\end{equation}

 The electric field phasor $\#E_{\,exc} (\#r_{\,a}) $ \textit{exciting} the sphere of material `a'
 is not merely $\#E_{\,inc} (\#r_{\,a}) $; instead,
\begin{equation} \l{ea1}
\#E_{\,exc}(\#r_{\,a}) =  \#E_{\,inc}(\#r_{\,a}) + \#E_{\,sca}^{b} (\#r_{\,a}),
\end{equation}
where   $\#E_{\,sca}^{b} $ represents the electric field  scattered by the sphere of material `b'. Likewise,
  there are two contributions to $\#E_{\,exc} ( \#r_{b}) $; i.e.,
\begin{equation} \l{ea2}
\#E_{\,exc}( \#r_{b}) =  \#E_{\,inc}( \#r_{b}) + \#E_{\,sca}^{a} ( \#r_{\,b}),
\end{equation}
where   $\#E_{\,sca}^{a} $ represents the electric field scattered by the sphere of material `a'.
 In the quasi-electrostatic regime, $ \#E_{\,inc}( \#r_{\,a}) \simeq  \#E_{\,inc}( \#r_{\,b}) \simeq  \#E_{\,inc}( \#0)$ and the scattered field phasors are given by \cite[Sec.~10.5]{Chen}
\begin{equation} \l{dipole_field}
\left.
\begin{array}{l}
\#E_{\,sca}^{a} (\#r_{\,b}) =
\displaystyle{
\frac{1}{4 \pi \eps^{\perp}_{out} d^3} \les \,  2   \, \#{\hat{d}} \,  \#{\hat{d}} -  \gamma \,
  \le
\=I - \#{\hat{d}} \, \#{\hat{d}} \ri
 \, \ris \. \#p_{\,a} } \vspace{6pt}
\\
\#E_{\,sca}^{b} (\#r_{\,a}) = \displaystyle{
\frac{1}{4 \pi \eps^{\perp}_{out} d^3} \les \,  2   \, \#{\hat{d}} \,  \#{\hat{d}} -  \gamma \,  \le
\=I - \#{\hat{d}} \, \#{\hat{d}} \ri  \, \ris  \. \#p_{\,b}  }
\end{array}
\right\}.
\end{equation}

The combination of Eqs.~\r{p_a_dipole}, \r{ea1}, \r{ea2}, and \r{dipole_field} yields
\begin{equation}
\left.
\begin{array}{l}
\displaystyle{\#p_{\,a} - \sigma_\perp  \, \=\alpha_{\, a/out} \. \les \, 2  \, \#{\hat{d}} \, \#{\hat{d}}   - \gamma  \le
\=I - \#{\hat{d}} \, \#{\hat{d}} \ri  \, \ris \. \#p_{\,b}  =
\frac{4}{3} \pi  s^3  \,    \=\alpha_{\, a/out} \. \#E_{\,inc}( \#0)} \vspace{6pt}
\\
\displaystyle{
\#p_{\,b} - \sigma_\perp  \, \=\alpha_{\, b/out} \. \les \,  2  \, \#{\hat{d}} \, \#{\hat{d}}   - \gamma
 \le
\=I - \#{\hat{d}} \, \#{\hat{d}} \ri
 \, \ris \. \#p_{\,a}  =
\frac{4}{3} \pi  s^3  \,    \=\alpha_{\, b/out} \. \#E_{\,inc}( \#0)}
\end{array}
\right\},
 \l{pa12einc}
\end{equation}
wherein the  parameter
\begin{equation} \l{sigma_def}
\sigma_\perp =
 \frac{s^3}{ 3 \eps^\perp_{out} d^3}\,.
\end{equation}
The pair of linear Eqs.~\r{pa12einc}
deliver the electric dipole moment
\begin{equation}  \#p_{\,\ell}  =
\frac{4}{3} \pi  s^3 \,  \={\tilde{\alpha}}_{\,\ell/out}   \.  \#E_{\,inc}( \#0), \qquad   \ell \in\lec a,b \ric, \l{pMEinc}
\end{equation}
 with
 the 3$\times$3 dyadic function
\begin{eqnarray} \l{alpha_uniaxial_a}
\={\tilde{\alpha}}_{\, \ell /out}  &=& \les \le \, 1 -  \sigma_{\perp}^2 \gamma^2 \alpha^\perp_{a/out} \alpha^\perp_{b/out}  \ri \le
\=I - \#{\hat{d}} \, \#{\hat{d}} \ri + \le 1 -  4 \sigma_{\perp}^2 \alpha^\parallel_{a/out} \alpha^\parallel_{b/out} \ri
\, \#{\hat{d}} \, \#{\hat{d}} \,  \ris^{-1} \nonumber \\ && \.
\les \le \alpha^\perp_{ \ell/out} - \sigma_{\perp} \gamma \alpha^\perp_{a/out} \alpha^\perp_{b/out}   \ri  \le
\=I - \#{\hat{d}} \, \#{\hat{d}} \ri +
\le \alpha^\parallel_{ \ell/out} + 2 \sigma_{\perp} \alpha^\parallel_{a/out} \alpha^\parallel_{b/out}  \ri \, \#{\hat{d}} \, \#{\hat{d}}  \, \ris.
\end{eqnarray}

The dyadic  $ \={\tilde{\alpha}}_{\, \ell/out} \ne \= {\alpha}_{\, \ell/out}$ is the polarizability density dyadic
of a monomer (sphere) of material `$\ell$' as a constituent of an isolated dimer immersed in the material labeled `out'.
The sum
\begin{equation}
\={\alpha}_{\, dimer /out}= \={\tilde{\alpha}}_{\, a /out}+\={\tilde{\alpha}}_{\, b /out}
\end{equation}
may be regarded as the polarizability density dyadic of the dimer, and the electric dipole moment
\begin{equation}
\l{eq18}
 \#p_{\,dimer} =  \#p_{\,a}+ \#p_{\,b}= \frac{4}{3} \pi  s^3\,\={\alpha}_{\, dimer /out} \.  \#E_{\,inc}( \#0)
 \end{equation}
characterizes the quasi-static scattering response of the dimer. Although a uniaxial object \cite{WLM1993},   the dimer
is different from   a rod or a needle because the volume entering the right side  of Eq.~\r{eq18} is that of a sphere but not of  a cylinder.

Next, let us turn to the electrically small sphere of
material `c'   immersed in a uniaxial dielectric material characterized by the permittivity
  dyadic $\=\eps_{\,out}$ given in  Eq.~\r{eps_out}.
 The sphere is centered at the  origin $\#r = \#0$. Suppose that the sphere is illuminated by a quasi-electrostatic  field $\#E_{\,inc} (\#r) $. The induced electric dipole moment is  given by \c{Jackson}
\begin{equation} \l{p_b_dipole}
\#p_{\,c} = \frac{4}{3} \pi  s^3 \=\alpha_{\,c/out} \.  \#E_{\,inc} (\#0),
\end{equation}
where the polarizability density dyadic $\=\alpha_{\,c/out}$ is defined  per Eq.~\r{a_polarizability_def} but with $\eps_{ \ell}$ therein replaced by $\eps_c$, and the components of $\=\alpha_{\,c/out}$ are written as
$\alpha^\perp_{c/out}$ and $\alpha^\parallel_{c/out}$ per Eq.~\r{D_alpha_form}${}_2$.

\subsection{Isotropic dielectric immersion material} \l{isotropic_out_sec}

Suppose that the immersion material is isotropic, i.e., $\=\eps_{\,out}=\eps_{\,out}\=I$. Then, the derivations
in \S\ref{uniaxial_out_sec} simply considerably. In particular, the depolarization dyadic $\=D_{\,out}$ reduces
to
 $\le3 i \omega   \eps_{out} \ri^{-1} \, \=I$,
while the
polarizability density dyadic $\=\alpha^{}_{\ell/out}$ reduces to
$ \alpha_{\ell/out} \=I $, where the
polarizability density scalar
\begin{equation} \l{alpha_a_scalar}
\alpha_{ \ell/out} = 3 \eps_{out} \le \frac{ \eps_{\ell} - \eps_{out} }{\eps_{ \ell} + 2 \eps_{out}}  \ri, \qquad \ell \in\lec a,b\ric.
\end{equation}
Consequently, we get
\begin{equation} \l{alpha_isotropic}
\left.
\begin{array}{l}
\={\tilde{\alpha}}_{\,a/out}  = \les \le \, 1 -  \sigma^2  \alpha_{a/out} \alpha_{b/out}  \ri \le
\=I - \#{\hat{d}} \, \#{\hat{d}} \ri + \le 1 -  4 \sigma^2 \alpha_{a/out} \alpha_{b/out} \ri
\, \#{\hat{d}} \, \#{\hat{d}} \,  \ris^{-1}  \vspace{6pt}  \\ \hspace{10mm} \.
\les \le 1  - \sigma   \alpha_{b/out}   \ri  \le
\=I - \#{\hat{d}} \, \#{\hat{d}} \ri +
\le 1 + 2 \sigma  \alpha_{b/out}  \ri \, \#{\hat{d}} \, \#{\hat{d}}  \, \ris \alpha_{a /out} \vspace{8pt} \\
\={\tilde{\alpha}}_{\,b/out}  = \les \le \, 1 -  \sigma^2  \alpha_{a/out} \alpha_{b/out}  \ri \le
\=I - \#{\hat{d}} \, \#{\hat{d}} \ri + \le 1 -  4 \sigma^2 \alpha_{a/out} \alpha_{b/out} \ri
\, \#{\hat{d}} \, \#{\hat{d}} \,  \ris^{-1}  \vspace{6pt}  \\ \hspace{10mm} \.
\les \le 1  - \sigma  \alpha_{a /out}   \ri  \le
\=I - \#{\hat{d}} \, \#{\hat{d}} \ri +
\le 1 + 2 \sigma  \alpha_{a /out}  \ri \, \#{\hat{d}} \, \#{\hat{d}}  \, \ris \alpha_{b/out}
\end{array}
\right\},
\end{equation}
wherein the  parameter
\begin{equation}
\sigma  =    \frac{s^3}{ 3 \eps_{out} d^3}\,.
\end{equation}
These simple expressions  are useful when the homogenization of a composite material
containing randomly oriented dimers is considered.

The special case of homodimers is noteworthy. Here $\eps_{a} = \eps_{b} $ and thus
 $\alpha_{a/out} = \alpha_{b/out}$.
 Hence,   $\={\tilde{\alpha}}_{\,a/out} = \={\tilde{\alpha}}_{\,b/out}  $ with
\begin{equation} \l{alpha_isotropic_homo}
\={\tilde{\alpha}}_{\,a/out} (\tilde{\sigma}) = \les \le \, 1 -  \tilde{\sigma}^2  \ri \le
\=I - \#{\hat{d}} \, \#{\hat{d}} \ri + \le 1 -  4 \tilde{\sigma}^2  \ri
\, \#{\hat{d}} \, \#{\hat{d}} \,  \ris^{-1}   \.
\les \le 1  - \tilde{\sigma}    \ri  \le
\=I - \#{\hat{d}} \, \#{\hat{d}} \ri +
\le 1 + 2 \tilde{\sigma }  \ri \, \#{\hat{d}} \, \#{\hat{d}}  \, \ris \alpha_{a/out} ,
\end{equation}
and the dimensionless scalar parameter
\begin{equation} \l{tsig_def}
\tilde{\sigma} = \le \frac{ \eps_{a } - \eps_{out} }{\eps_{a } + 2 \eps_{out}}   \ri \frac{s^3}{d^3}.
\end{equation}

The electric dipole moment of an electrically small sphere of material `c' is given as in Eq.~\r{p_b_dipole} with $\=\alpha_{\,c/out}=\alpha_{\,c/out} \, \=I$, where
\begin{equation} \l{alpha_b_scalar}
\alpha_{c/out} = 3 \eps_{out} \le \frac{ \eps_{c} - \eps_{out} }{\eps_{c} + 2 \eps_{out}}  \ri.
\end{equation}

 \section{Identically oriented dimers} \l{Sec_UOD}

If  all dimers  have the same orientation then the  HCM is a uniaxial dielectric material \cite{Chen, EAB} with
its  distinguished axis parallel to the fixed unit vector  $\#{\hat{d}} $. That is, $\=\eps_{HCM}$ has the form given in Eq.~\r{e_HCM}.

\subsection{Bruggeman formalism}

Particles of all component materials  are assumed as being immersed in the HCM itself,
in the Bruggeman formalism \cite{L96,ML_PiO}. Thus,  the expressions presented in \S\ref{uniaxial_out_sec} are appropriate here with
 the subscript `Br' replacing the subscript `out'.
The Bruggeman formalism rests on upon the assumption that  the  electric dipole moments arising from the
electrically small spheres of the component materials, weighted by volume fraction, sum to the zero vector \cite{RL2005};
i.e., 
\begin{eqnarray}
&&   f_a\,   \#p_{\,dimer }    + f_c \, \#p_{\,c}   =
f_a \left( \#p_{\,a }  +  \#p_{\,b } \right) + f_c \, \#p_{\,c}   = \#0\,. \l{Br_oriented}
\end{eqnarray}
Upon combining Eqs.~\r{pMEinc}--\r{p_b_dipole} with Eq.~\r{Br_oriented},
the  dyadic  equation
\begin{eqnarray}
\=0&=&
\l{dyadic_Br_eqn}
f_a \, \={\alpha}_{\,dimer/Br}
+ f_c \, \=\alpha_{\,c/Br}
\\[6pt]
\nonumber
&=&
f_a \, \={\tilde{\alpha}}_{\,a/Br}   + f_b \, \={\tilde{\alpha}}_{\,b/Br}
+ f_c \, \=\alpha_{\,c/Br}
\\[6pt]
&=& f_a
\les \le \, 1 -  \sigma^2 \gamma^2 \alpha^\perp_{a/Br} \alpha^\perp_{b/Br}  \ri \le
\=I - \#{\hat{d}} \, \#{\hat{d}} \ri + \le 1 -  4 \sigma^2 \alpha^\parallel_{a/Br} \alpha^\parallel_{b/Br} \ri
\, \#{\hat{d}} \, \#{\hat{d}} \,  \ris^{-1} \nonumber
\\ &&
\nonumber
\.
\left[  \le \alpha^\perp_{a/Br} + \alpha^\perp_{b/Br} - 2 \sigma \gamma \alpha^\perp_{a/Br} \alpha^\perp_{b/Br}   \ri
  \le
\=I - \#{\hat{d}} \, \#{\hat{d}} \ri  \right.
\\
\nonumber
&&\left.
+
\le \alpha^\parallel_{a/Br} + \alpha^\parallel_{b/Br}  + 4 \sigma \alpha^\parallel_{a/Br} \alpha^\parallel_{b/Br}  \ri \, \#{\hat{d}} \, \#{\hat{d}} \,
 \right]
 \\ &&
 + f_c \les \, \alpha^{\perp}_{c/Br} \le \=I - \#{\hat{d}} \, \#{\hat{d}} \ri + \alpha^{\parallel}_{c/Br}\, \#{\hat{d}} \, \#{\hat{d}} \, \ris\,
  \l{Br_dyadic_comp}
\end{eqnarray}
emerges.

Due to the uniaxial symmetry, Eq.~\r{Br_dyadic_comp} represents
two coupled  scalar equations with  $\eps^{\perp}_{Br}$ and $\eps^{\parallel}_{Br}$ as the two unknowns, which must
be obtained by  numerical methods. The following Jacobi scheme may be used for this purpose \c{Jacobi}. First, let us notice that
$\={\tilde{\alpha}}_{\,a,b/Br}$, as defined in Eq.~\r{alpha_uniaxial_a} (with the subscript `Br' replacing the subscript `out'), may be written as
\begin{equation} \l{aM_eqn}
\left.
\begin{array}{l}
\={\tilde{\alpha}}_{\,a/Br}   = \=M_{\,b} \. \=\alpha_{\,a/Br} \vspace{6pt} \\
\={\tilde{\alpha}}_{\,b/Br}   = \=M_{\,a} \. \=\alpha_{\,b/Br}
\end{array}
\right\},
\end{equation}
wherein the 3$\times$3 dyadic
\begin{eqnarray}
\=M_{\,\ell}
&=& \lec \ \=I - \sigma^2 \les \, \gamma^2 \le
\=I - \#{\hat{d}} \, \#{\hat{d}} \ri + 4 \, \#{\hat{d}} \, \#{\hat{d}} \ris \. \=\alpha_{\,a/Br} \. \=\alpha_{\,b/Br}\ric^{-1} \nonumber \\ &&
\. \lec \=I + \sigma \les \, - \gamma \le
\=I - \#{\hat{d}} \, \#{\hat{d}} \ri + 2 \, \#{\hat{d}} \, \#{\hat{d}} \ris \.  \=\alpha_{\,\ell/Br} \ric, \qquad
 \ell \in\lec a,b \ric.
\end{eqnarray}
Second, notice that $\=\alpha_{\,a,b,c/Br}$, as defined in Eq.~\r{a_polarizability_def} (with the subscript `Br' replacing the subscript `out'), may be written as
\begin{equation} \l{a_polarizability_def_P}
\=\alpha^{}_{\,\ell/Br} =  \left(\eps_{\ell } \=I
- \=\eps_{\,Br} \right)\. \=P_{\,\ell} ,  \qquad   \ell \in\lec a, b, c \ric,
\end{equation}
wherein
 the 3$\times$3 dyadic
\begin{eqnarray}
\=P_{\,\ell} = \les \=I + i \omega \=D_{\,Br} \.
\le \eps_{\ell } \=I - \=\eps_{\,Br} \ri
\ris^{-1}.
\end{eqnarray}
Hence, after using Eqs.~\r{aM_eqn} and \r{a_polarizability_def_P},   Eq.~\r{dyadic_Br_eqn} may expressed as
\begin{equation} \l{dyadic_Br_eqn_2}
f_a \les \, \=M_{\,b} \. \le \eps_{a} \=I - \=\eps_{\,Br}  \ri\. \=P_{\,a}
+  \=M_{\,a} \. \le \eps_{b} \=I - \=\eps_{\,Br}\ri \. \=P_{\,b}   \ris
+ f_c \le \eps_{c}  - \=\eps_{\,Br}\ri \. \=P_{\,c} = \=0\,.
\end{equation}
The Bruggeman estimate $\=\eps_{\,Br}$ may be extracted from Eq.~\r{dyadic_Br_eqn_2} by the iterative scheme represented by
\begin{equation}
\=\eps_{\,Br} = \mathcal{T} \lec \=\eps_{\,Br} \ric,
\end{equation}
where the action of the dyadic operator $\mathcal{T}$ is given by
\begin{eqnarray}
\mathcal{T} \lec \=\eps_{\,Br} \ric &=& \les f_a \le \=M_{\,b} \. \=P_{\,a} + \=M_{\,a} \. \=P_{\,b} \ri
+ f_c \=P_{\,c} \ris^{-1} \nonumber \\ && \.
\les f_a \le \eps_{a} \=M_{\,b} \. \=P_{\,a} + \eps_{b} \=M_{\,a} \. \=P_{\,b} \ri
+ f_c \eps_c \=P_{\,c} \ris.
\end{eqnarray}

\subsection{Maxwell Garnett formalism}

Particles of component materials `a' and `b' are viewed as immersed in component material `c', in the Maxwell Garnett formalism \cite{L96}.
Thus, the expressions presented in \S\ref{isotropic_out_sec} are appropriate here with the subscript `c' replacing the subscript `out'.
 The electric dipole moments of spheres of component material `c' are not relevant to this formalism,
 the HCM essentially arising as  a perturbation of the component material `c' by the addition of a relatively small amount of component materials `a' and `b'. Consequently, results of the Maxwell Garnett formalism, as applied here, are  strictly appropriate only for  $ f_a \lessapprox 0.15$.

The Maxwell Garnett estimate of $\=\eps_{\,HCM}$ is obtained explicitly as \c{WLM97}
\begin{eqnarray}
\=\eps_{\,MG} &=&
 \eps_{c}  \, \=I + f_a \, \lec \, \={\tilde{\alpha}}_{\,a/c}   \. \les \, \=I -  \frac{ 2 f_a}{3 \eps_c}
 \, \={\tilde{\alpha}}_{\,a/c}   \ris^{-1} +
 \={\tilde{\alpha}}_{\,b/c}   \. \les \, \=I -  \frac{2 f_a}{3 \eps_c}
\, \={\tilde{\alpha}}_{\,b/c}   \ris^{-1} \ric \,,
\end{eqnarray}
with the polarizability density dyadics $\={\tilde{\alpha}}_{\,a,b/c}$  as given in Eqs.~\r{alpha_isotropic}.

\section{Randomly oriented dimers} \l{Sec_ROD}

If  the
 dimers  are randomly oriented, the HCM is isotropic. Accordingly, the electrically small spheres of component materials `a' and `b' (and `c' in the case of the Bruggeman formalism)
 should be regarded in this case as being immersed in an isotropic dielectric material and
the expressions presented in \S\ref{isotropic_out_sec}
are appropriate.

Orientationally averaged electric dipole moments  are defined as
\begin{equation}
\langle \, \#p_{\, \ell} \rangle = \frac{1}{4 \pi}  \int^{2 \pi}_{\phi = 0} \int^{\pi}_{\theta = 0} \#p_{\, \ell} \, \sin \theta \, d \theta \, d \phi, \qquad \ell \in\lec a,b,c\ric.
\end{equation}
 The  orientationally averaged electric dipole moments for component materials `a' and `b' may be expressed as
\begin{equation} \l{ave_p_int}
\langle \, \#p_{\,  \ell} \rangle =  \frac{4 \pi}{3} s^3  \le \frac{1}{4 \pi} \int^{2 \pi}_{\phi = 0} \int^{\pi}_{\theta = 0} \={\tilde{\alpha}}_{\,\ell/out}   \, \sin \theta \, d \theta \, d \phi \ri \. \#E_{\, inc}( \#0), \qquad \ell \in\lec a,b\ric.
\end{equation}
Herein the quantity in parenthesis represents the orientational average of the polarizability density dyadic
$\={\tilde{\alpha}}_{\ell/out}$; for later use, this is written as
\begin{equation}
\langle \, \={\tilde{\alpha}}_{\,\ell/out}   \rangle =  {\tilde{\alpha}}^{ave}_{\, \ell/out}    \, \=I\,,
\qquad \ell \in\lec a,b\ric.
\end{equation}
Let us also note that
\begin{equation}
\left.\begin{array}{c}
\langle \, \#p_{\,  dimer} \rangle = \langle \, \#p_{\,  a} \rangle + \langle \, \#p_{\, b} \rangle
\\[6pt]
\langle \, \={\alpha}_{\,dimer/out}   \rangle = \langle \, \={\tilde{\alpha}}_{\,a/out}   \rangle
+ \langle \, \={\tilde{\alpha}}_{\,b/out}   \rangle
\\[6pt]
{\alpha}^{ave}_{\, dimer/out}  = {\tilde{\alpha}}^{ave}_{\, a/out} +{\tilde{\alpha}}^{ave}_{\,b/out}
\end{array}\right\}\,.
\end{equation}

Upon evaluating the integrals on the right side of Eq.~\r{ave_p_int}, the following result is delivered:
\begin{equation} \l{ave_p_int2}
\left.
\begin{array}{l}
\displaystyle{
\langle \, \#p_{\,a} \rangle = \frac{4}{9} \pi  s^3 \alpha_{a/out}
  \les \frac{2 \le 1 -  \sigma \alpha_{b/out} \ri}{1-  \sigma^2 \alpha_{a/out} \alpha_{b/out}
 } +
  \frac{1+ 2 \sigma \alpha_{b/out}}{1- 4 \sigma^2 \alpha_{a/out} \alpha_{b/out} } \,\ris  \,\#E_{\,inc}( \#0) }
  \vspace{8pt}
   \\
  \displaystyle{ \langle \, \#p_{\,b} \rangle = \frac{4}{9} \pi  s^3 \alpha_{b/out}
  \les \frac{2 \le 1 -  \sigma \alpha_{a/out} \ri}{1-  \sigma^2 \alpha_{a/out} \alpha_{b/out}
 } +
  \frac{1+ 2 \sigma \alpha_{a/out}}{1- 4 \sigma^2 \alpha_{a/out} \alpha_{b/out} } \,\ris  \,\#E_{\,inc}( \#0)}
   \end{array}
   \right\}.
\end{equation}
In the special case of homodimers, $\#p_{\,a} = \#p_{\,b}$ and the corresponding orientationally averaged  electric dipole moment is given as
\begin{equation} \l{ave_p_int2_homo}
\langle \#p_{a } \rangle = \langle \#p_{b } \rangle = \frac{4}{3} \pi  s^3 \alpha_{a /out}
  \les \frac{\tilde{\sigma} - 1}{\le \tilde{\sigma} + 1 \ri \le 2 \tilde{\sigma} -1  \ri}
  \,\ris  \,\#E_{\,inc}( \#0).
\end{equation}

Since spheres of  material `c'  have no directional dependency, the orientational average of the associated electric dipole moment  is simply $\#p_{c}$ itself; i.e.,
\begin{equation} \l{pb_ave}
\langle \#p_{\,c} \rangle =  \frac{4}{3} \pi  s^3 \alpha_{c/out} \, \#E_{\,inc} (\#0).
\end{equation}

\subsection{Bruggeman formalism}

In the Bruggeman formalism, particles of all component materials  are assumed as being immersed in the HCM itself.
Thus, the expressions presented in \S\ref{isotropic_out_sec} can be used here with
the subscript `out' replaced by the subscript `Br'.
The Bruggeman formalism dictates that
\begin{eqnarray}
&& 
 f_a \langle \#p_{\,dimer} \rangle   + f_c \langle \#p_{\,c} \rangle=
f_a \left( \langle \#p_{\,a} \rangle +  \langle \#p_{\,b} \rangle\right)
  + f_c \langle \#p_{\,c} \rangle  = \#0\,. \l{Br_ave}
\end{eqnarray}
Upon combining Eqs.~\r{ave_p_int2} and \r{pb_ave} with Eq.~\r{Br_ave},
the corresponding scalar Bruggeman equation emerges as 
\begin{eqnarray} \l{scalar_Br}
&&
f_a \left\{\frac{\alpha_{a/Br}}{3} \les \frac{2 \le 1 -  \sigma \alpha_{b/Br} \ri}{1-  \sigma^2 \alpha_{a/Br} \alpha_{b/Br}
 } +
  \frac{1+ 2 \sigma \alpha_{b/Br}}{1- 4 \sigma^2 \alpha_{a/Br} \alpha_{b/Br} } \,\ris \right.
\nonumber \\ &&
+ \left.   \frac{\alpha_{b/Br}}{3}
\les \frac{2 \le 1 -  \sigma \alpha_{a/Br} \ri}{1-  \sigma^2 \alpha_{a/Br} \alpha_{b/Br}
 } +
  \frac{1+ 2 \sigma \alpha_{a/Br}}{1- 4 \sigma^2 \alpha_{a/Br} \alpha_{b/Br} } \,\ris\right\}
+ f_c \alpha_{c/Br}
= 0.
\end{eqnarray}
After using Eqs.~\r{sigma_def}, \r{alpha_a_scalar}, and \r{alpha_b_scalar} to
 substitute for $\sigma$, $\alpha_{a,b/Br}$, and $\alpha_{c/Br}$,   respectively,
Eq.~\r{scalar_Br} may be recast as a quintic polynomial in $\eps_{Br}$. (In the case of homodimers, this polynomial reduces to a cubic polynomial in $\eps_{Br}$). A Jacobi numerical scheme \c{Jacobi} can be used to extract $\eps_{Br}$ from Eq.~\r{scalar_Br}. That is,
the solution may be found using the iterative scheme represented by
\begin{equation}
\eps_{Br} = \mathcal{S} \lec \eps_{Br} \ric,
\end{equation}
where the action of the scalar operator $\mathcal{S}$ is given by
\begin{equation}
\mathcal{S} \lec \eps_{Br} \ric =  \frac{ \displaystyle{ \frac{m_b \, \eps_a}{\eps_a + 2 \eps_{Br}}
+\frac{m_a \, \eps_b}{\eps_b + 2 \eps_{Br}} +\frac{f_c \,\eps_c}{\eps_c + 2 \eps_{Br}} }}
{ \displaystyle{
\frac{m_b }{\eps_a + 2 \eps_{Br}}
+\frac{m_a }{\eps_b + 2 \eps_{Br}} +\frac{f_c }{\eps_c + 2 \eps_{Br}}
}} \, ,
\end{equation}
with the scalar parameters
\begin{equation}
\left.
\begin{array}{l}
m_a = \displaystyle{\frac{f_a}{3} \les \frac{2 \le 1 -  \sigma \alpha_{a/Br} \ri}{1-  \sigma^2 \alpha_{a/Br} \alpha_{b/Br}
 } +
  \frac{1+ 2 \sigma \alpha_{a/Br}}{1- 4 \sigma^2 \alpha_{a/Br} \alpha_{b/Br} } \,\ris} \vspace{8pt} \\
m_b = \displaystyle{\frac{f_a}{3} \les \frac{2 \le 1 -  \sigma \alpha_{b/Br} \ri}{1-  \sigma^2 \alpha_{a/Br} \alpha_{b/Br}
 } +
  \frac{1+ 2 \sigma \alpha_{b/Br}}{1- 4 \sigma^2 \alpha_{a/Br} \alpha_{b/Br} } \,\ris}
\end{array}
\right\}.
\end{equation}

\subsection{Maxwell Garnett formalism}

In the Maxwell Garnett formalism,
particles of component materials `a' and `b' are viewed as immersed in component material `c'.
Accordingly, here
 the expressions presented in \S\ref{isotropic_out_sec} are used with the subscript `c' replacing the subscript `out'.
 The Maxwell Garnett estimate of $\eps_{HCM}$ is given explicitly by
 \begin{equation}
 \eps_{MG} = \eps_c \lec 1 + 3 f_a \, \les \, \frac{  {\tilde{\alpha}}^{ave}_{\,a/c}  } { 3 \, \eps_c - 2 f_a
 \, {\tilde{\alpha}}^{ave}_{\,a/c}   } +
 \frac{  {\tilde{\alpha}}^{ave}_{\,b/c}   } { 3 \,\eps_c  - 2 f_a
 \, {\tilde{\alpha}}^{ave}_{\,b/c}   }
  \ris  \ric \,,
 \end{equation}
where
\begin{equation}
\left.
\begin{array}{l}
\displaystyle{
 {\tilde{\alpha}}^{ave}_{\,a/c}  =  \frac{1}{3}  \alpha_{a/c}
  \les \frac{2 \le 1 -  \sigma \alpha_{b/c} \ri}{1-  \sigma^2 \alpha_{a/c} \alpha_{b/c}
 } +
  \frac{1+ 2 \sigma \alpha_{b/c}}{1- 4 \sigma^2 \alpha_{a/c} \alpha_{b/c} } \,\ris} \vspace{8pt} \\
  \displaystyle{
 {\tilde{\alpha}}^{ave}_{\,b/c}  =  \frac{1}{3}  \alpha_{b/c}
  \les \frac{2 \le 1 -  \sigma \alpha_{a/c} \ri}{1-  \sigma^2 \alpha_{a/c} \alpha_{b/c}
 } +
  \frac{1+ 2 \sigma \alpha_{a/c}}{1- 4 \sigma^2 \alpha_{a/c} \alpha_{b/c} } \,\ris}
\end{array}
\right\}\,.
\end{equation}

\section{Numerical results} \l{Num_sec}

Let us now present   representative numerical results based on the theoretical results established in \S\ref{Homog_prelim}--\S\ref{Sec_ROD}. 
Although the range $f_a\in[0,0.5]$ may appear appropriate
at first glance, the maximum value of $f_a$ must be less than $0.5$. This is because no sphere of material `c' must  be allowed to occupy the region between the two spheres constituting a dimer. Nevertheless, we have provided the Bruggeman estimates  for $f_a\in[0,0.5]$, because the upper limit of $f_a$ will have to be decided experimentally for a
specific composite material. The Maxwell Garnett formalism is appropriate only for dilute composite materials, and we have restricted the presentation of
 the Maxwell Garnett estimates to $ f_a \in[0, 0.15]$. 

\subsection{Dielectric--dielectric dimers} \l{dd_numerical}
Suppose, first, that both component materials `a' and `b' are nondissipative dielectric materials, specified by the permittivities
$\eps_a = 2 \epso$ and $\eps_b \in \le \epso, 10\epso \ri $. The dielectric--dielectric dimers which arise from the combination of
component materials `a' and `b'
 are randomly dispersed along with component material `c'   specified by the permittivity $ \eps_c  = \le 14 + 4 i \ri \epso$.

\subsubsection{Randomly oriented dimers}

The real and imaginary parts of $\eps_{MG}/\epso$
and $\eps_{Br}/\epso$ are plotted against $ \eps_b/\epso $ and $f_a $ in Fig.~\ref{fig1}, for the case where the
dimers are randomly oriented and $d= 2s$.
 The real part of $\eps_{MG}$ decreases approximately linearly as $f_a$ increases, with its rate of decrease being greatest at the lowest  values of $\eps_b$. Furthermore, the
real part of $\eps_{MG}$ increases approximately linearly as $\eps_b$ increases, with its rate of increase being greatest at the largest values of $f_a$. The imaginary part of $\eps_{MG}$ decreases approximately linearly as $f_a$ increases, and this trend is almost independent of the value of $\eps_b$. For the range  $0 < f_a \lessapprox 0.15$, the real and imaginary parts of $\eps_{Br}$ are very similar, both qualitatively and quantitatively, to the real and imaginary parts of $\eps_{MG}$. For  $f_a \gtrapprox 0.15$, both the real and imaginary parts of $\eps_{Br}$ exhibit a more nonlinear dependency on $f_a$ than they do at lower values of $f_a$.

The issue of the influence of the dimer separation distance upon  $\eps_{HCM}$ is addressed via Fig.~\ref{fig2},
wherein
the real and imaginary parts of $\eps_{MG}/\epso$
and $\eps_{Br}/\epso$ are plotted against  $  d / s $ for   $ \eps_b  = \epso $ (green, solid curves), $2 \epso$ (blue, dashed curves), and $3 \epso$ (red, broken dashed curves), when $f_a = 0.15$. The influence of  $d$ on the real and imaginary parts of $\eps_{HCM}$, for both the Bruggeman and Maxwell Garnett estimates, decays rapidly as $d$ increases. Indeed,  both estimates of $\eps_{HCM}$ are practically independent of $d$ for  $d > 4 s$,   these estimates being essentially the same as those that would be obtained through the homogenization of three independent component materials `a', `b', and `c' with no dimeric interaction between the spheres of component materials `a' and `b'.
The change in the real part of $\eps_{MG}$ as $d$ increases from zero to $4 s$ is approximately 0.2$\%$, whereas the corresponding change in the imaginary part of $\eps_{MG}$ is approximately 0.4$\%$. The Bruggeman estimates of $\eps_{HCM}$ are somewhat more sensitive to changes in $d$, the real part of $\eps_{Br}$ changing by approximately 1.6$\%$ whereas
the imaginary part of $\eps_{Br}$ changing by approximately 2.8$\%$ as $d$ increases from zero to $4 s$.

\subsubsection{Identically oriented dimers}

Qualitatively, the Bruggeman and Maxwell Garnett estimates of $\eps^{\parallel}_{HCM}$ and $\eps^{\perp}_{HCM}$ for
composite materials containing identically oriented dimers
have dependencies similar to the  estimates of $\eps_{HCM}$ presented in Figs.~\ref{fig1} and \ref{fig2} for composite materials containing randomly oriented dimers.  This becomes evident from the plots of the averages
$\le \eps^{\parallel}_{MG} + \eps^{\perp}_{MG} \ri /2\epso$
and $\le \eps^{\parallel}_{Br} + \eps^{\perp}_{Br} \ri /2\epso$ with respect to $\eps_b/\epso$ and $f_a$ in Fig.~\ref{figX}.
To the naked eye, the plots in Figs.~\ref{fig1} and \ref{figX} are almost indistinguishable.

However, there are significant quantitative differences between the estimates of $\eps^{\parallel}_{HCM}$ and $\eps^{\perp}_{HCM}$.
In Fig.~\ref{fig3}, the real and imaginary parts
 of the differences $\le \eps^{\parallel}_{MG} - \eps^{\perp}_{MG} \ri /\epso$
and $\le \eps^{\parallel}_{Br} - \eps^{\perp}_{Br} \ri/\epso$ are
plotted against $ \eps_b/\epso $ and $f_a $, when $d= 2s$. Both the real and imaginary parts of  the difference $ \eps^{\parallel}_{MG} - \eps^{\perp}_{MG}$ increase approximately linearly as $f_a$ increases and decrease approximately linearly as $\eps_b$ increases. Thus, the greatest degree of anisotropy is
predicted by the Maxwell Garnett formalism when $f_a$ is largest and $\eps_b$ is smallest.

In the range  $0 < f_a \lessapprox 0.15$, both the real and imaginary parts of
 the difference $ \eps^{\parallel}_{Br} - \eps^{\perp}_{Br}$
 are qualitatively similar to the corresponding real and imaginary parts of $ \eps^{\parallel}_{MG} - \eps^{\perp}_{MG}$.
However,  at larger values of $f_a$, both the real and imaginary parts of $ \eps^{\parallel}_{Br} - \eps^{\perp}_{Br}$  exhibit  strong nonlinear dependencies on $f_a$. The greatest degree of anisotropy   is predicted
by the Bruggeman formalism to be in the vicinity of $f_a \approx 0.2$  with $\eps_b = \epso$. Broadly, over the parameter ranges considered,  the degree of anisotropy estimated by the Maxwell Garnett formalism is slightly larger than that estimated by the  Bruggeman formalism.

The degree of anisotropy exhibited by the HCM, as estimated by the Bruggeman and Maxwell Garnett formalisms, decays rapidly as the  separation distance $d$ in the dimer increases. This is illustrated in Fig.~\ref{fig4} wherein
the real and imaginary parts of
the differences $\le \eps^{\parallel}_{MG} - \eps^{\perp}_{MG} \ri /\epso$
and $\le \eps^{\parallel}_{Br} - \eps^{\perp}_{Br} \ri/\epso$
are plotted against  $  d / s $ for  $ \eps_b  = \epso $ (green, solid curves), $2 \epso$ (blue, dashed curves), and $3 \epso$ (red, broken dashed curves). Here $f_a = 0.15$. The graphs for $\eps^{\parallel}_{Br} - \eps^{\perp}_{Br}$ and $\eps^{\parallel}_{MG} - \eps^{\perp}_{MG}$ in Fig.~\ref{fig4} are qualitatively similar,  with the  Maxwell Garnett estimates being slightly larger than the Bruggeman estimates at each value of $d$ and $\eps_b$. The degree of anisotropy, as estimated by both formalisms, falls most rapidly for the smallest value of $\eps_b$. Furthermore,  the degree of anisotropy, as estimated by  both formalisms, vanishes almost entirely at $d = 4 s$.

\subsection{Metal--dielectric dimers} \l{md_numerical}

Next, suppose that
 component material `a' is a metal. For definiteness,  this metal is taken to be silver as
   characterized by the size-dependent permittivity \c{BH,Kreibig}
\begin{equation} \l{Drude_formula_2}
\eps_{Ag} (s) = \epso \les 1 - \frac{\omega^2_p}{\omega^2 + i \, \omega \,
\le \gamma_{Ag} + \displaystyle{ \frac{3 v_F}{4s} } \ri } \ris.
\end{equation}
Herein,
$v_F = 1.4 \times 10^{6}$m~$\mbox{s}^{-1}$  is the
electron speed at the Fermi surface, $\gamma_{Ag} = 10^{14}$~s${}^{-1}$  is the relaxation rate, and $\omega_p = 1.38 \times 10^{16}$~rad~s$^{-1}$ is
 the plasma frequency. The angular frequency is  $\omega = 2 \pi \co / \lambdao$, with the free--space wavelength chosen to be  $\lambdao = 650$~nm.
 As in \S\ref{dd_numerical}, component material `b' is a nondissipative dielectric material specified by the permittivity
 $\eps_b \in \le \epso, 10 \epso\ri$. The metal--dielectric dimers which arise from the combination of
component materials `a' and `b'
 are randomly mixed with component material `c' which is specified by the permittivity $ \eps_c  = \le 14 + 4 i \ri \epso$
 for all results presented here.

\subsubsection{Randomly oriented dimers}

\begin{figure}[!h]
\centering \psfull
 \epsfig{file=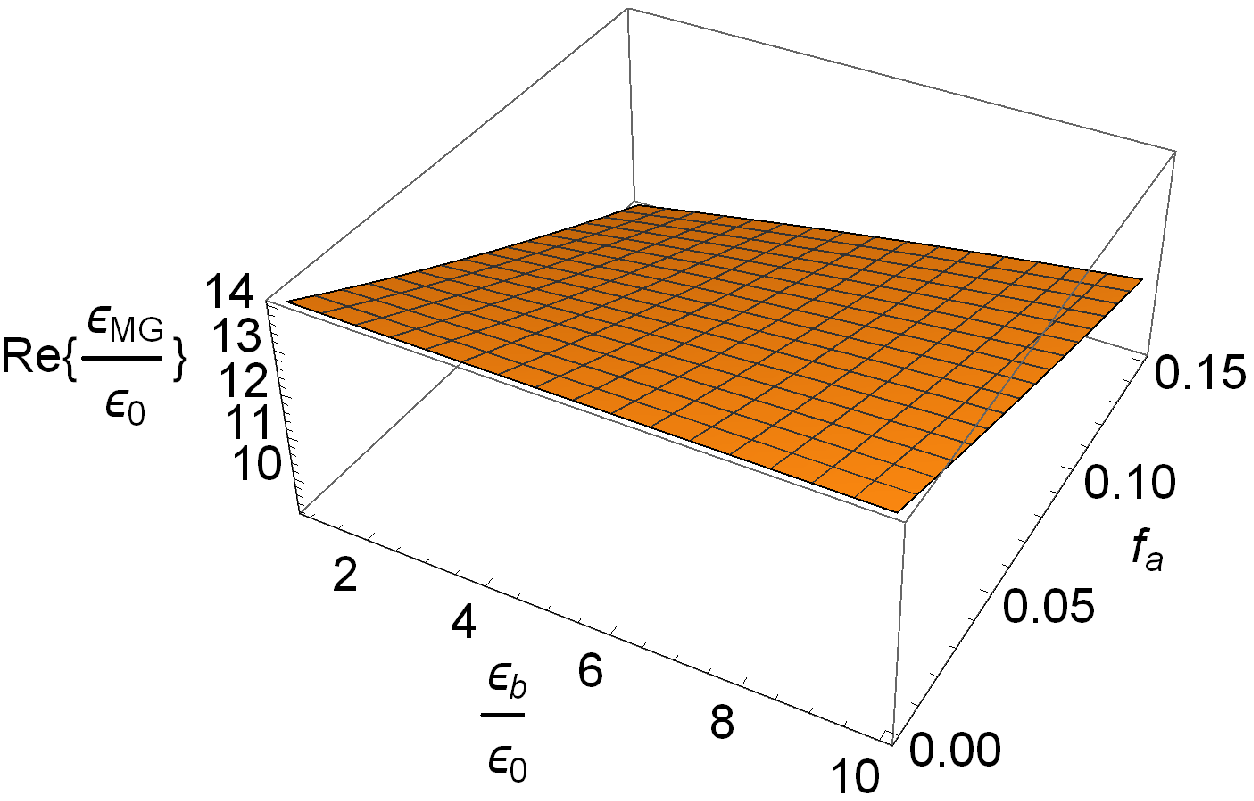,width=3.1in} \hfill
\epsfig{file=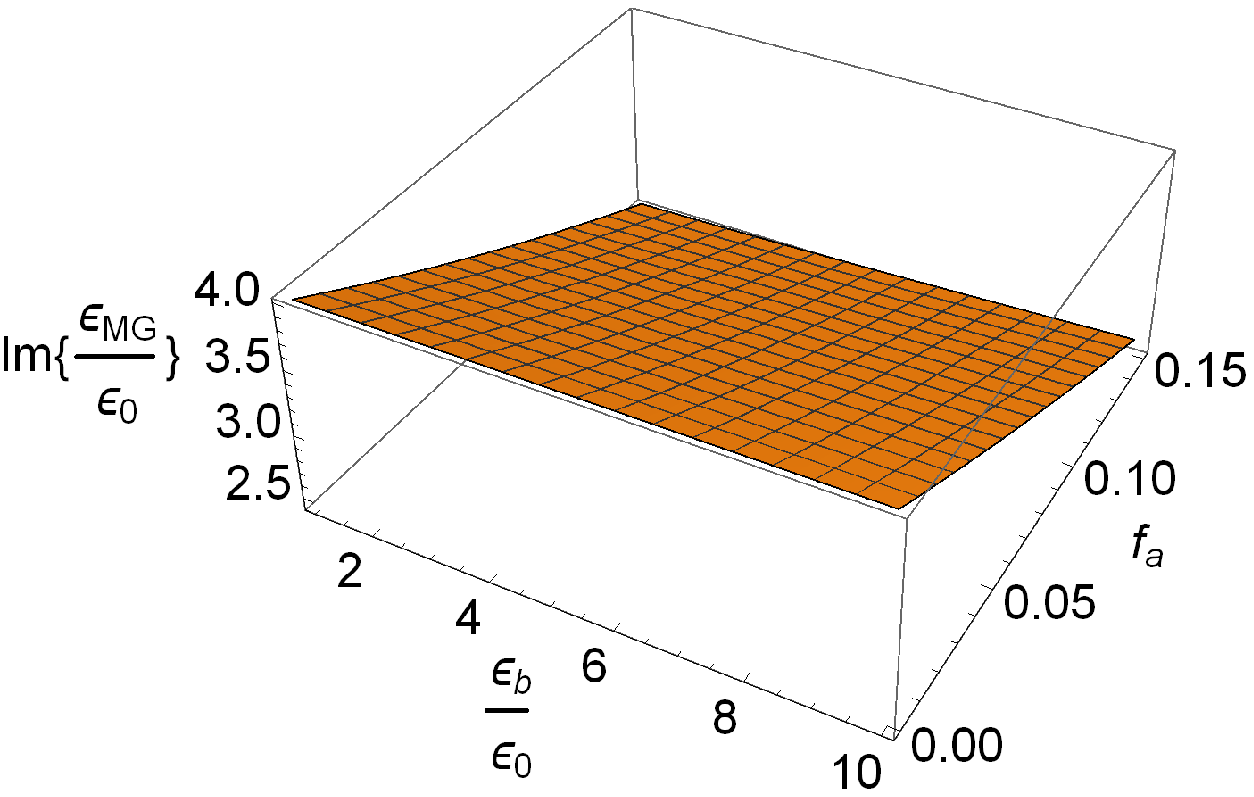,width=3.1in} \vspace{15mm} \\
\epsfig{file=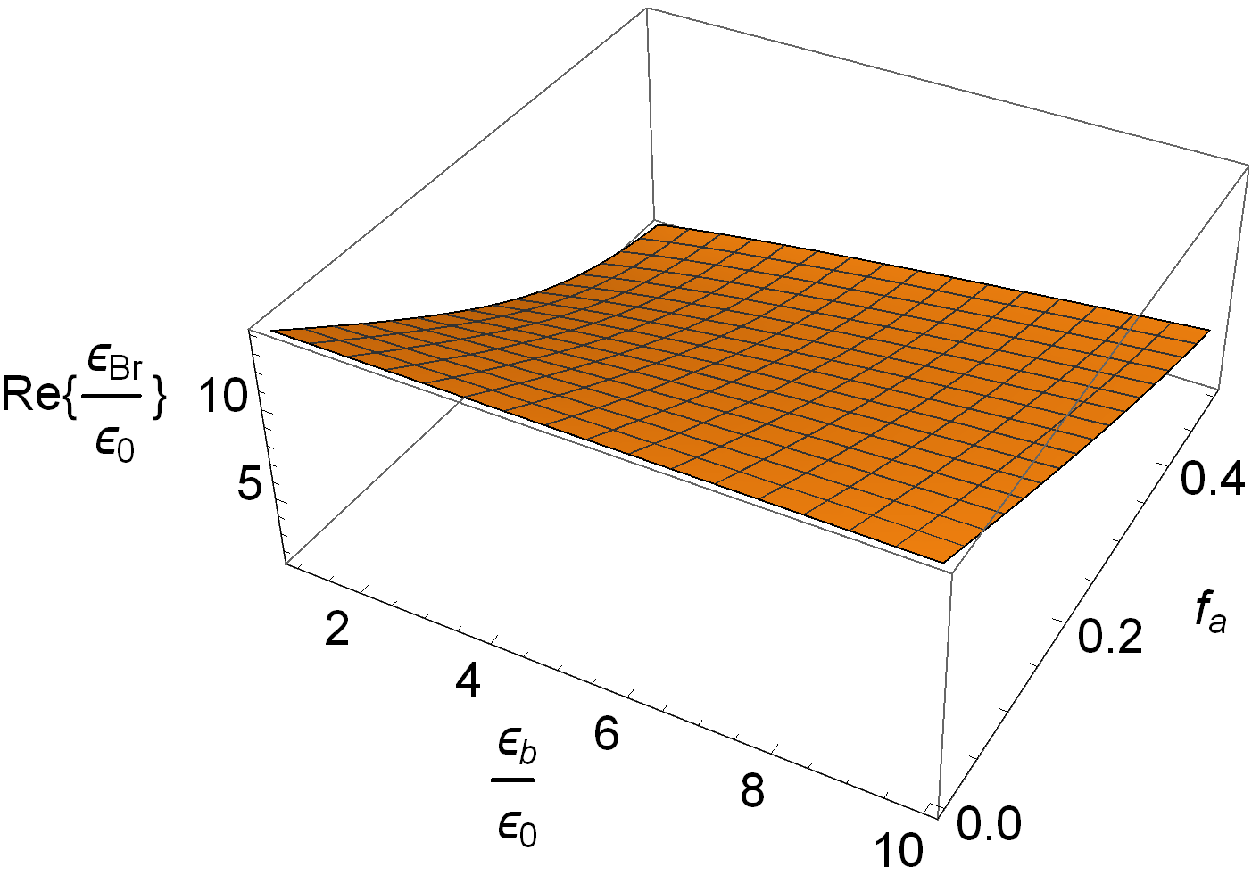,width=3.1in} \hfill
\epsfig{file=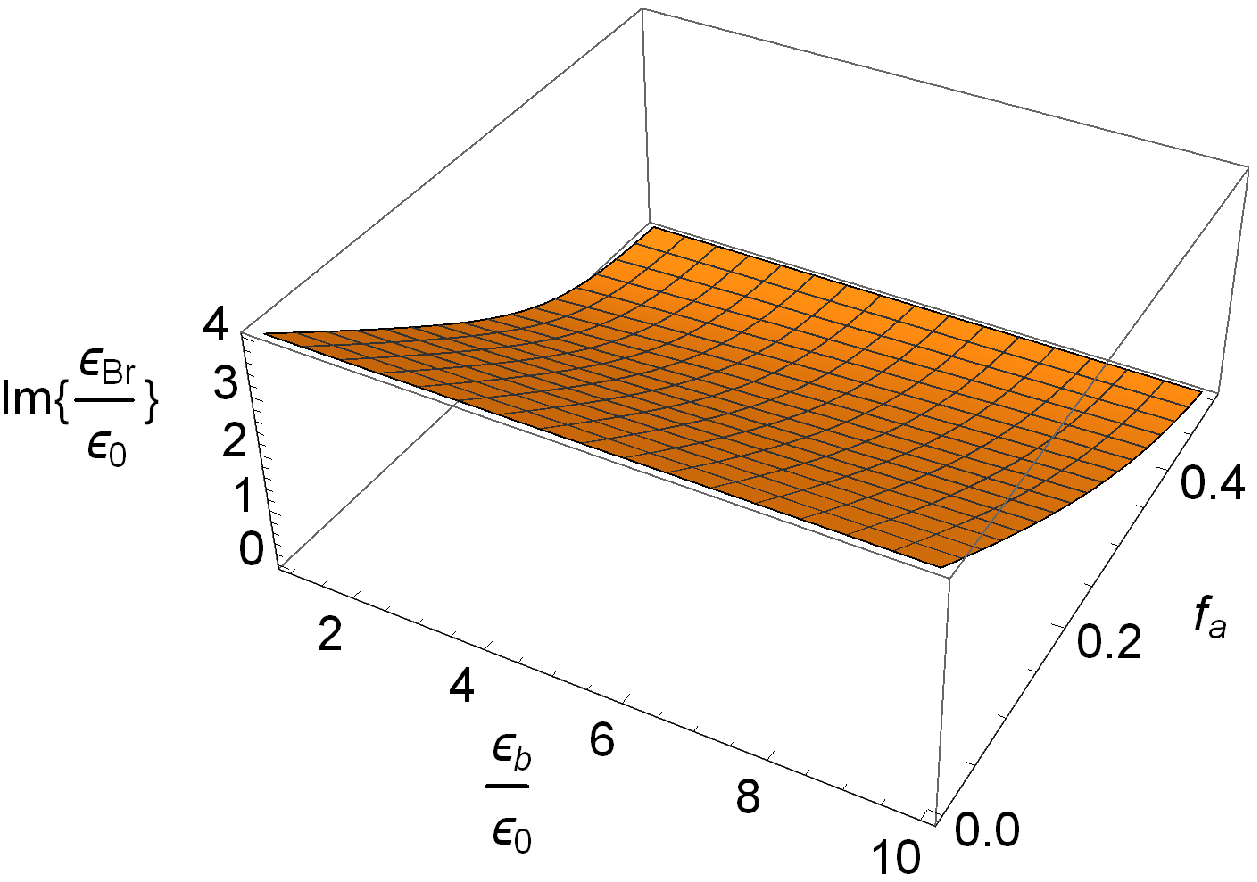,width=3.1in}
\caption{The real and imaginary parts of $\eps_{MG}/\epso$
and $\eps_{Br}/\epso$ plotted against $ \eps_b/\epso $ and $f_a $ for the case where the electrically small
spheres of component materials `a' and `b' combine to form dielectric--dielectric dimers with $\eps_a = 2 \epso$ and $\eps_b \in \le 1, 10 \ri \epso$.
 Component material `c' is specified by
$ \eps_c  = \le 14 + 4 i \ri \epso$. The dimers are randomly oriented and $d= 2s$.  }
\label{fig1}
\end{figure}

For composite materials containing randomly oriented dimers,
the real and imaginary parts of $\eps_{MG}/\epso$
and $\eps_{Br}/\epso$ are plotted in Fig.~\ref{fig5} against $ \eps_b/\epso $ and $f_a $ for $d= 2s$ and $s = 5$ nm. Thus, $\eps_a = \le -21.4 + 2.4 i \ri \epso$  by virtue of Eq.~\r{Drude_formula_2}.
The graphs of the real parts of $\eps_{MG}$ and $\eps_{Br}$ in Fig.~\ref{fig5} are qualitatively similar to the corresponding graphs in Fig.~\ref{fig1}
for dielectric--dielectric dimers. In contrast, graphs of the imaginary parts of $\eps_{MG}$ and $\eps_{Br}$ in Fig.~\ref{fig5} are
rather different to the corresponding graphs in Fig.~\ref{fig1}, both qualitatively and quantitatively. On the whole, the imaginary parts of $\eps_{MG}$ and $\eps_{Br}$
 are substantially larger in Fig.~\ref{fig5} than they are in Fig.~\ref{fig1}. Furthermore,
 the imaginary parts of $\eps_{MG}$ and $\eps_{Br}$
  in Fig.~\ref{fig5} are substantially more nonlinear with respect to increasing $f_a$ than are the corresponding quantities  in Fig.~\ref{fig1}.  Both the real and imaginary parts of $\eps_{MG}$ and $\eps_{Br}$ are qualitatively similar in Fig.~\ref{fig5} in  the range  $0 < f_a \lessapprox 0.15$. However, across this range, the quantitative differences between the estimates $\eps_{MG}$ and $\eps_{Br}$ are substantially larger than the corresponding differences presented in Fig.~\ref{fig1}, and these differences between $\eps_{MG}$ and $\eps_{Br}$ grow as $f_a$ increases from zero.

Qualitatively, the influence of the dimer separation distance $d$ upon  $\eps_{HCM}$ for the metal--dielectric dimer HCM
is similar to that for the dielectric--dielectric dimer HCM. This may be appreciated by comparing Fig.~\ref{fig2} with Fig.~\ref{fig6}. In Fig.~\ref{fig6},
the real and imaginary parts of $\eps_{MG}/\epso$
and $\eps_{Br}/\epso$ are plotted against  $  d / s $ for  $ \eps_b  = \epso $ (green, solid curves), $2 \epso$ (blue, dashed curves), and $3 \epso$ (red, broken dashed curves), for the metal--dielectric dimer HCM, with
 $f_a = 0.02$ and $s= 5$ nm. As is the case for dielectric--dielectric dimers  in Fig.~\ref{fig2},  the estimates of $\eps_{HCM}$ for metal--dielectric dimers in Fig.~\ref{fig6} are practically independent of $d$ for $d > 4 s $. The magnitudes of the relative changes in   $\eps_{HCM}$
 as $d$ increases from zero to $4 s$ in Fig.~\ref{fig6} are similar to those in Fig.~\ref{fig2}, with the Bruggeman estimates being somewhat more sensitive than the Maxwell Garnett estimates to changes in $d$.

The effects  of the size of the metal particles which make up component material `a' are delineated in Fig.~\ref{fig7}. Herein
the real and imaginary parts of $\eps_{MG}/\epso$
and $\eps_{Br}/\epso$ are plotted against  $ s $ for  $ \eps_b  = \epso $ (green, solid curves), $2 \epso$ (blue, dashed curves), and $3 \epso$ (red, broken dashed curves), with $f_a = 0.02$ and $d = 2s$.
The influence of $s$ on the real and imaginary parts of $\eps_{HCM}$, for both the Bruggeman and Maxwell Garnett estimates, decays rapidly as $s$ increases. Indeed,  the estimates of $\eps_{HCM}$ vary little as  $s$ increases beyond $20$~nm.
The change in the real part of $\eps_{MG}$ as $s$ increases from 5 to $20$ nm is approximately 2.6$\%$, whereas the corresponding change in the imaginary part of $\eps_{MG}$ is approximately 1.0$\%$. The real part of $\eps_{Br}$ changes by approximately 1.2$\%$ whereas
the imaginary part of $\eps_{Br}$ changes by approximately 1.8$\%$, as $s$ increases from 5 to $20$ nm. Most strikingly,
the real and imaginary parts of $\eps_{MG}$, as well as the real part of $\eps_{Br}$, uniformly decrease as $s$ increases from
5 to 20 nm whilst the imaginary part of $\eps_{Br}$ uniformly increases as $s$ increases.

\subsubsection{Identically oriented dimers}

The estimates of $\eps^{\parallel}_{HCM}$ and $\eps^{\perp}_{HCM}$
yielded by the Bruggeman and Maxwell Garnett formalisms for composite materials containing identically oriented dimers
exhibit characteristics that are qualitatively similar to those displayed in Figs.~\ref{fig5}--\ref{fig7} by
the corresponding estimates of $\eps_{HCM}$   for randomly oriented dimers. However,  significant quantitative differences arise between the estimates of $\eps^{\parallel}_{HCM}$ and $\eps^{\perp}_{HCM}$.
The real and imaginary parts
 of the differences $\le \eps^{\parallel}_{MG} - \eps^{\perp}_{MG} \ri /\epso$
and $\le \eps^{\parallel}_{Br} - \eps^{\perp}_{Br} \ri/\epso$ are
plotted in  Fig.~\ref{fig8} against $ \eps_b/\epso $ and $f_a $, for $d= 2s$ and $s = 5$~nm.
Across the range  $0 < f_a \lessapprox 0.15$,   the differences
 $ \eps^{\parallel}_{HCM} - \eps^{\perp}_{HCM} $
estimated by the Maxwell Garnett and
  Bruggeman formalisms are  qualitatively similar. However,  there are quantitative differences between $ \eps^{\parallel}_{Br} - \eps^{\perp}_{Br} $ and $ \eps^{\parallel}_{MG} - \eps^{\perp}_{MG} $, and these increase in magnitude as $f_a$ increases.
 The greatest degree of anisotropy is estimated by the Bruggeman formalism to exist when  both $f_a$  and $\eps_b $ are
 maximum. In contrast, the greatest degree of anisotropy is estimated by the Maxwell Garnett formalism to exist when  $f_a$  is maximum but $\eps_b $ is minimum.

The influences of the dimer separation distance $d$ and the metal sphere radius $s$ upon the anisotropy of the HCM are delineated in Fig.~\ref{fig9}. Here the real and imaginary parts
 of the differences $\le \eps^{\parallel}_{MG} - \eps^{\perp}_{MG} \ri /\epso$
and $\le \eps^{\parallel}_{Br} - \eps^{\perp}_{Br} \ri/\epso$ are
plotted against $ s $~(in nm) and $d/s $, for $f_a = 0.02$ and $d = 2s$. While the real part of $\eps^{\parallel}_{MG} - \eps^{\perp}_{MG} $
decreases uniformly as $d$ increases from $2s$ to $4s$,  this quantity varies very little as $s$ increases from 5~nm to 20~nm.
The imaginary part of $\eps^{\parallel}_{MG} - \eps^{\perp}_{MG} $
increases uniformly as $d$ increases from $2s$ to $4s$; in contrast, the imaginary part of $\eps^{\parallel}_{MG} - \eps^{\perp}_{MG} $ varies only marginally as $s$ increases from 5~nm to 20~nm. The graphs for the real and imaginary parts of $\eps^{\parallel}_{Br} - \eps^{\perp}_{Br} $ are both qualitatively and quantitatively similar to the corresponding graphs for
$\eps^{\parallel}_{MG} - \eps^{\perp}_{MG} $.

\section{Discussion} \l{Dis_sec}

The Bruggeman and Maxwell Garnett formalisms have been established in the preceding sections
for the homogenization of composite materials containing randomly oriented and identically oriented dimers. The representative numerical results presented in \S\ref{dd_numerical}
for the case of dielectric--dielectric dimers demonstrate close agreement between the estimates of the HCM constitutive parameters delivered by the
Bruggeman and Maxwell Garnett formalisms for both randomly oriented and identically oriented dimers. The Bruggeman formalism is advantageous over the Maxwell Garnett formalism insofar as the former is appropriate for arbitrary dimer volume fractions whereas the latter is appropriate only for low dimer volume fractions. On the other hand, the Maxwell Garnett formalism is relatively straightforward to implement numerically as its estimates are provided as explicit formulas, in contrast to the Bruggman formalism whose numerical implementation typically involves the iterative extraction of estimates from  implicit formulas.

The case of metal--dielectric dimers should generally be approached with caution. If attention is restricted to parameter regimes involving low dimer volume fractions and moderate degrees of dissipation then, as demonstrated in \S\ref{md_numerical}, the Bruggeman and Maxwell Garnett formalisms deliver estimates of the HCM constitutive parameters which are in broad agreement.
However, at higher dimer volume fractions, substantial qualitative and quantitative differences emerge between the estimates provided by the two  formalisms, and these differences are exacerbated by anisotropy in the case of identically oriented dimers.

The difficulties that arise for metal--dielectric dimers at larger values of the dimer volume fraction essentially stem from the fact the real part of $\eps_a$ is negative  while the real parts of $\eps_{b,c}$  are  positive.  In the absence of substantial degrees of dissipation, homogenization for such parameter regimes  can be  problematic for conventional formalisms, especially at mid-range values of  volume fractions \c{ML_Bruggeman,M_JNP,Bro1,Bro2}. For examples,  in these parameter regimes the Bruggeman estimates may violate the Hashin--Shtrikman bounds, and the Maxwell Garnett estimates may exhibit very large resonances (with respect to varying volume fraction). These issues effect both passive and active HCMs \c{Limit_active}, for both forward and inverse homogenization formalisms \c{Limit_inverse}, and also impose limitations on the Bergman--Milton bounds \c{Limit_BM}.

The mathematical origin of these problematic parameter regimes may be appreciated most readily by considering the simplest case, namely that of randomly oriented homodimers. The corresponding expression for the
orientationally averaged  electric dipole moment is provided in Eq.~\r{ave_p_int2_homo}. The polarizability scalar $\alpha_{a/out}$ therein becomes infinitely large in magnitude in the limit $\eps_a \to - 2 \eps_{out}$. This eventuality~---~which is sometimes referred to as a Fr\"ohlich mode \c{BH}~---~may result in singular or  highly resonant behavior in the estimates of HCM permittivity. There is further scope for singular behavior which is solely attributable to the dimer interaction: the denominator term in  Eq.~\r{ave_p_int2_homo} is null valued at $\tilde{\sigma} \in\lec -1,1/2\ric$. By using the definition of $\tilde{\sigma}$ provided in Eq.~\r{tsig_def} with $d = 2 s$, these singularities arise in the limits $\eps_a \to - (17/7) \eps_{out}$ and
$\eps_a \to - 3 \eps_{out}$, respectively. For strictly nondissipative materials, the possibility of $\eps_a/\eps_{out} \in \lec -2 ,- (17/7) , -3 \ric  $
 can only arise if either $\eps_a \eps_b < 0 $ or $\eps_a \eps_c < 0 $. Thus, by extrapolation, it may anticipated that regimes
 in which  the real part of $\eps_a$ is negative  while the real parts of at least one of  $\eps_{b}$ or $\eps_c$  is  positive  may well be problematic. However, as demonstrated in  \S\ref{md_numerical}, provided that only low dimer volume fractions are considered and
  there is a moderate degree of dissipation, the problems of  singular or highly resonant behavior may not arise.

\begin{figure}[!h]
\centering \psfull
 \epsfig{file=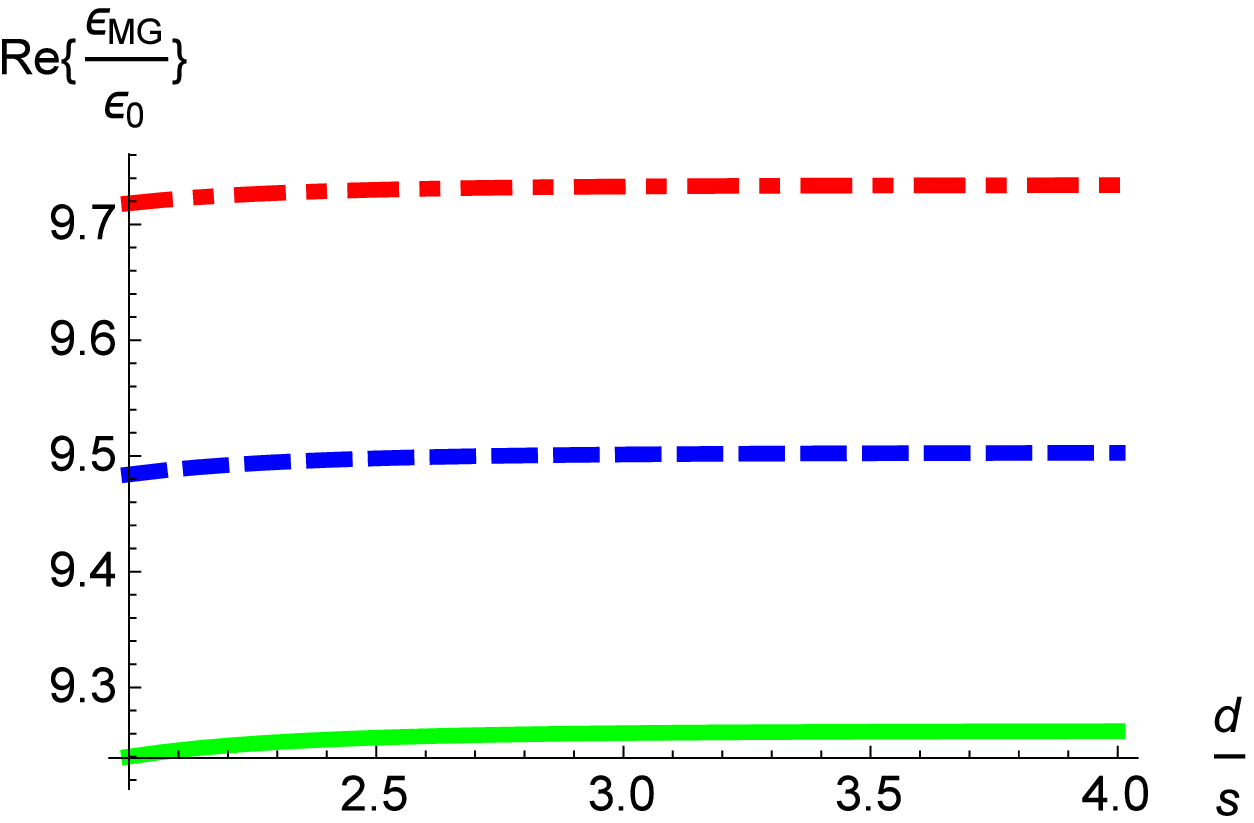,width=3.1in} \hfill
\epsfig{file=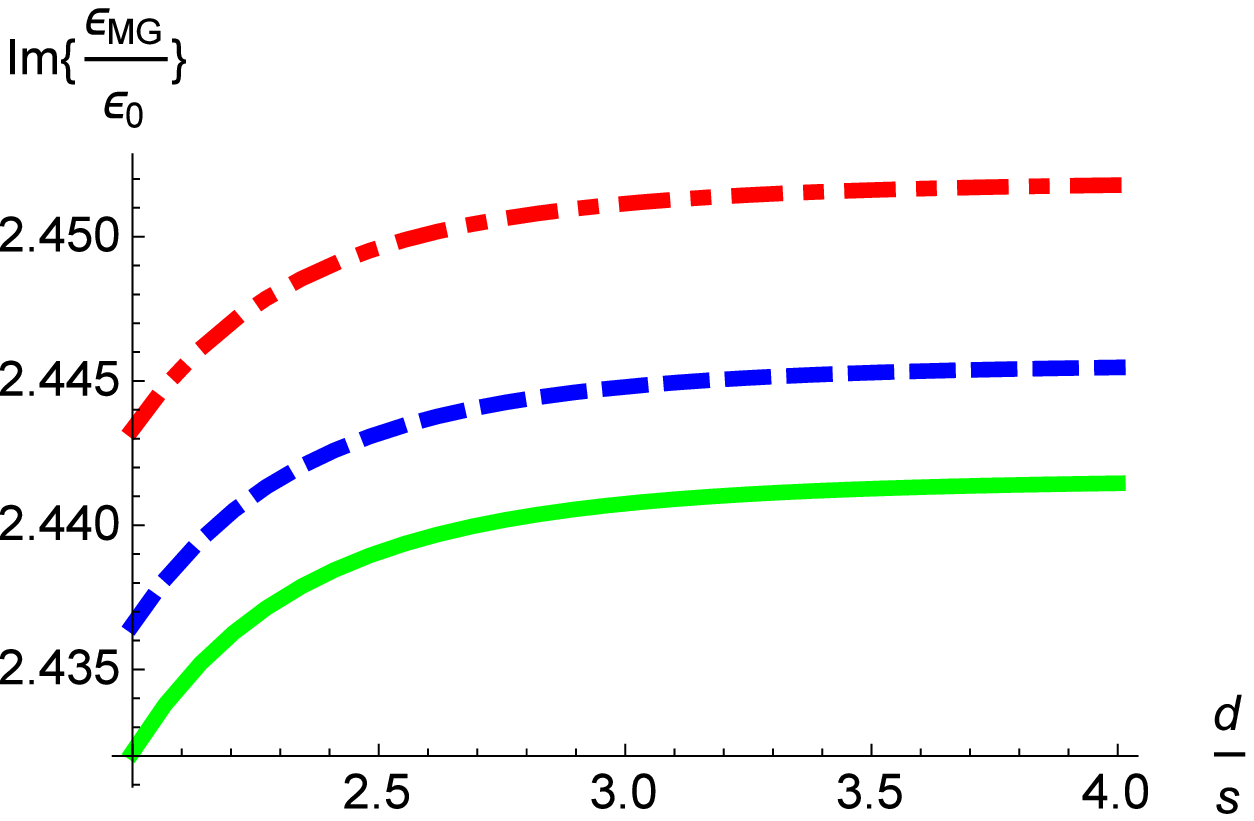,width=3.1in} \vspace{15mm} \\
\epsfig{file=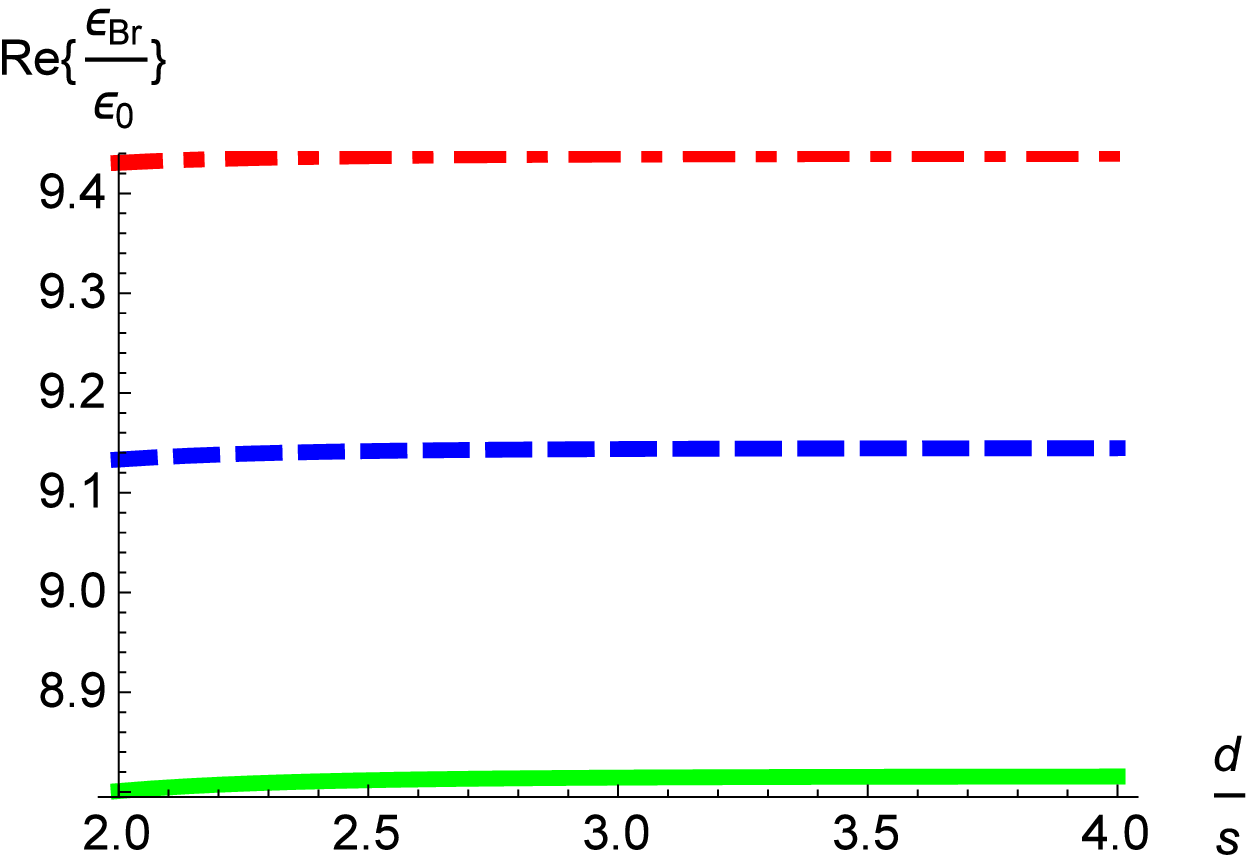,width=3.1in} \hfill
\epsfig{file=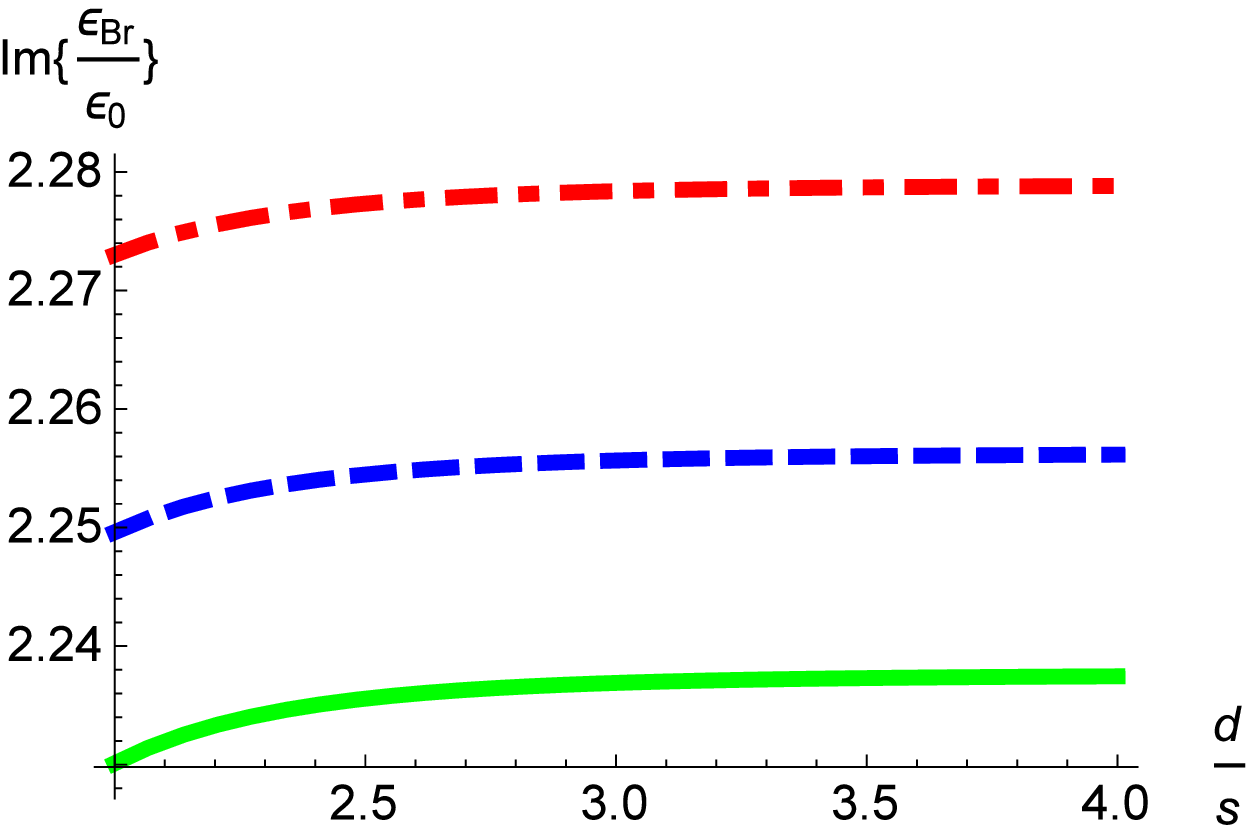,width=3.1in}
\caption{As Fig.~\ref{fig1} except that
the real and imaginary parts of $\eps_{MG}/\epso$
and $\eps_{Br}/\epso$ are plotted against  $  d / s $ for  $\eps_a = 2 \epso$ and $ \eps_b  = \epso $ (green, solid curves), $2 \epso$ (blue, dashed curves), and $3 \epso$ (red, broken dashed curves). Here $f_a = 0.15$.
 }
\label{fig2}
\end{figure}

\begin{figure}[!h]
\centering \psfull
 \epsfig{file=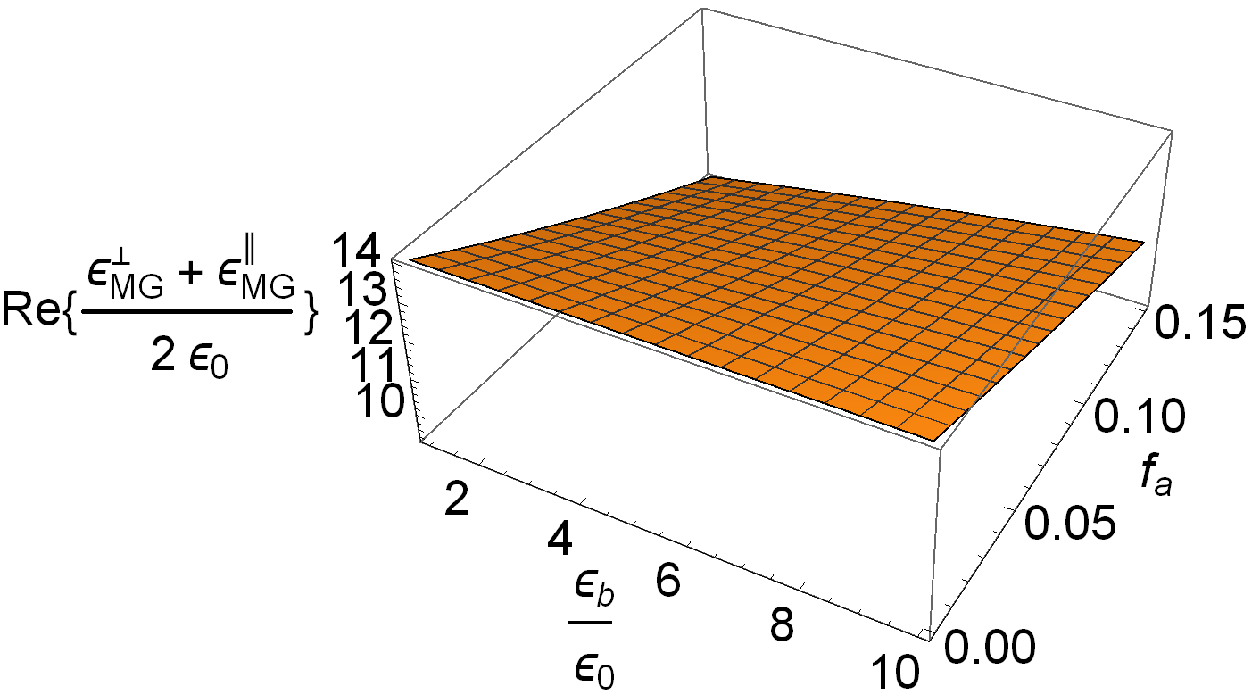,width=3.1in} \hfill
\epsfig{file=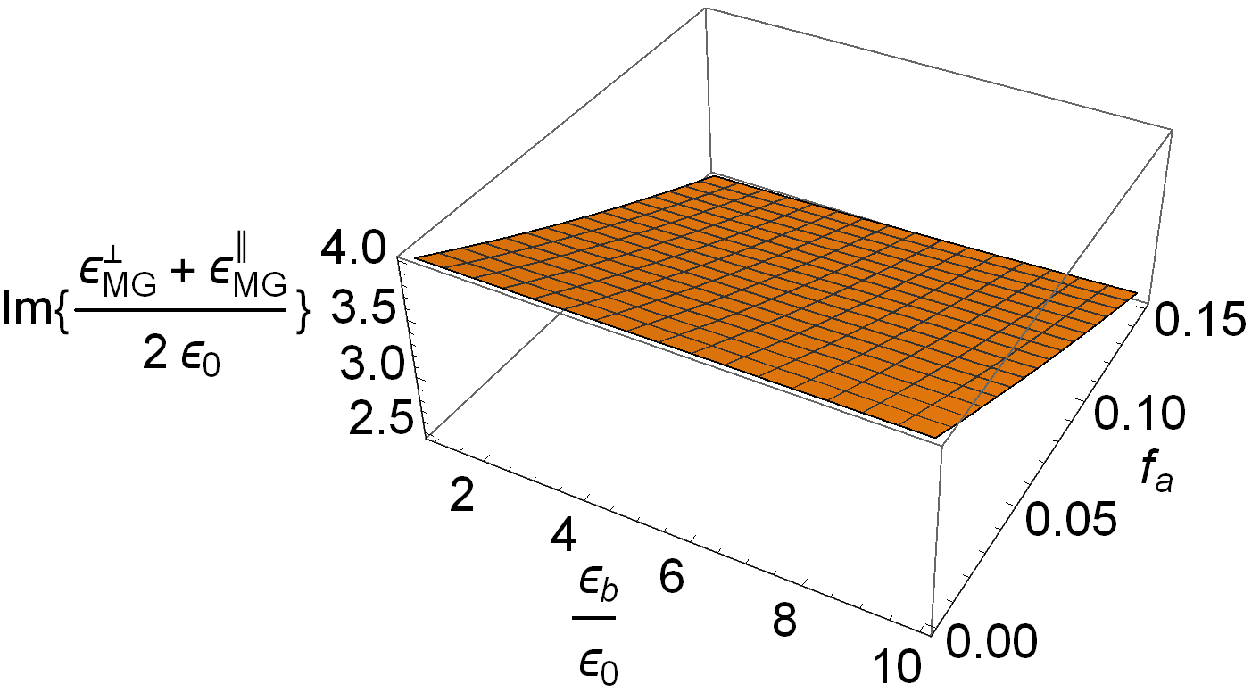,width=3.1in} \vspace{5mm} \\
\epsfig{file=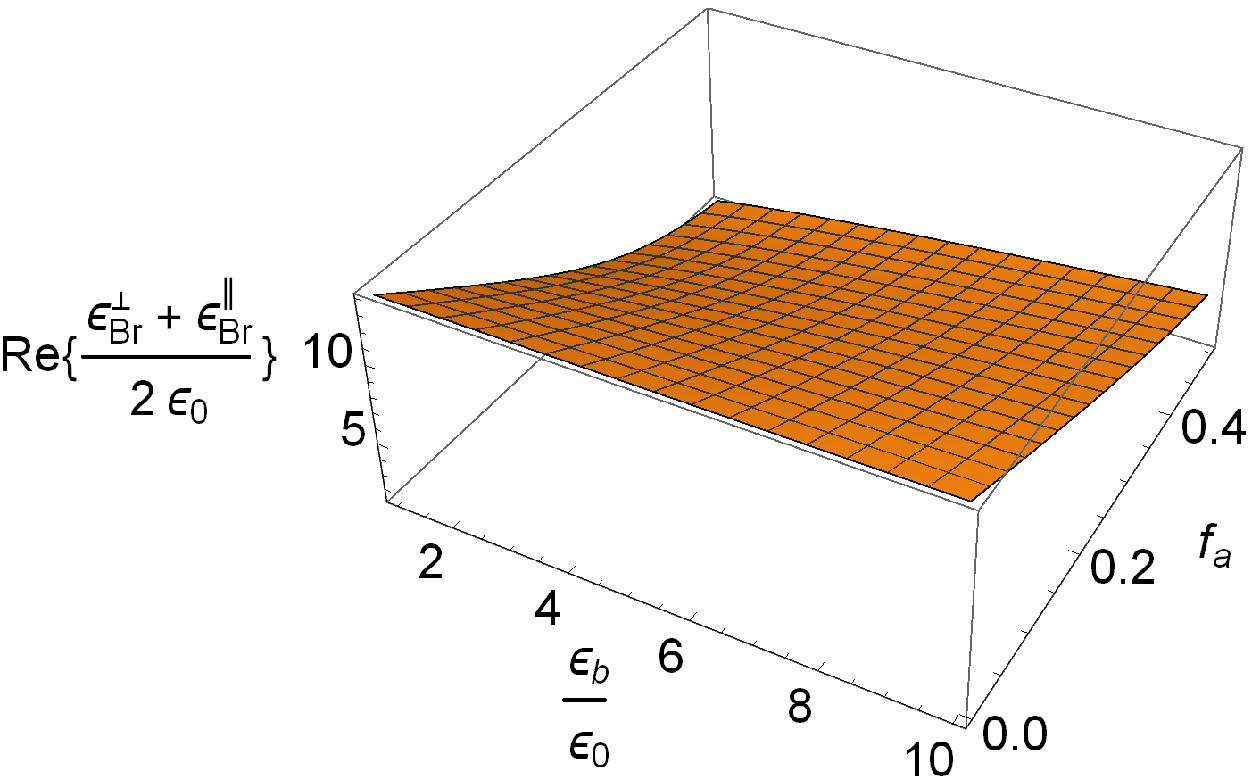,width=3.1in} \hfill
\epsfig{file=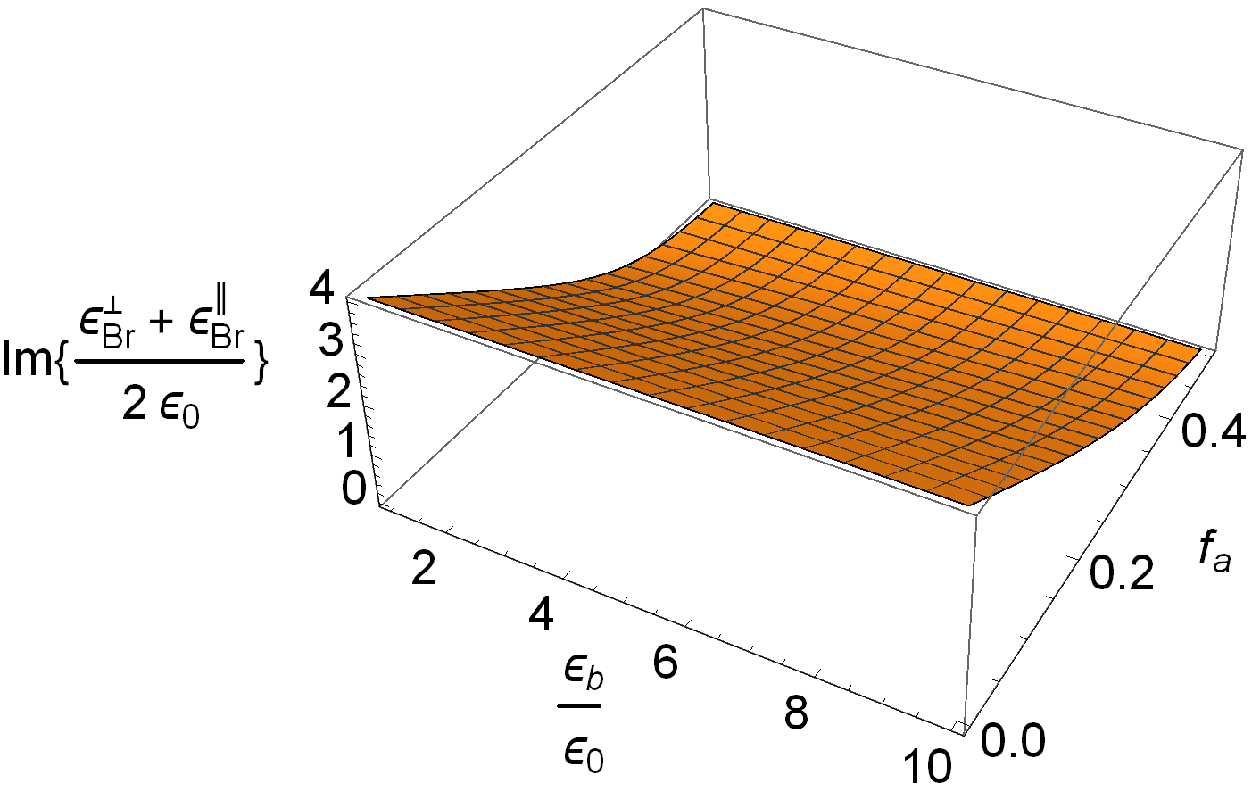,width=3.1in}
\caption{As Fig.~\ref{fig1} except that the dimers are identically oriented and the real and imaginary parts of the averages $\le \eps^{\parallel}_{MG} + \eps^{\perp}_{MG} \ri /2\epso$
and $\le \eps^{\parallel}_{Br} + \eps^{\perp}_{Br} \ri /2\epso$ are
plotted.
 }
\label{figX}
\end{figure}

\begin{figure}[!h]
\centering \psfull
 \epsfig{file=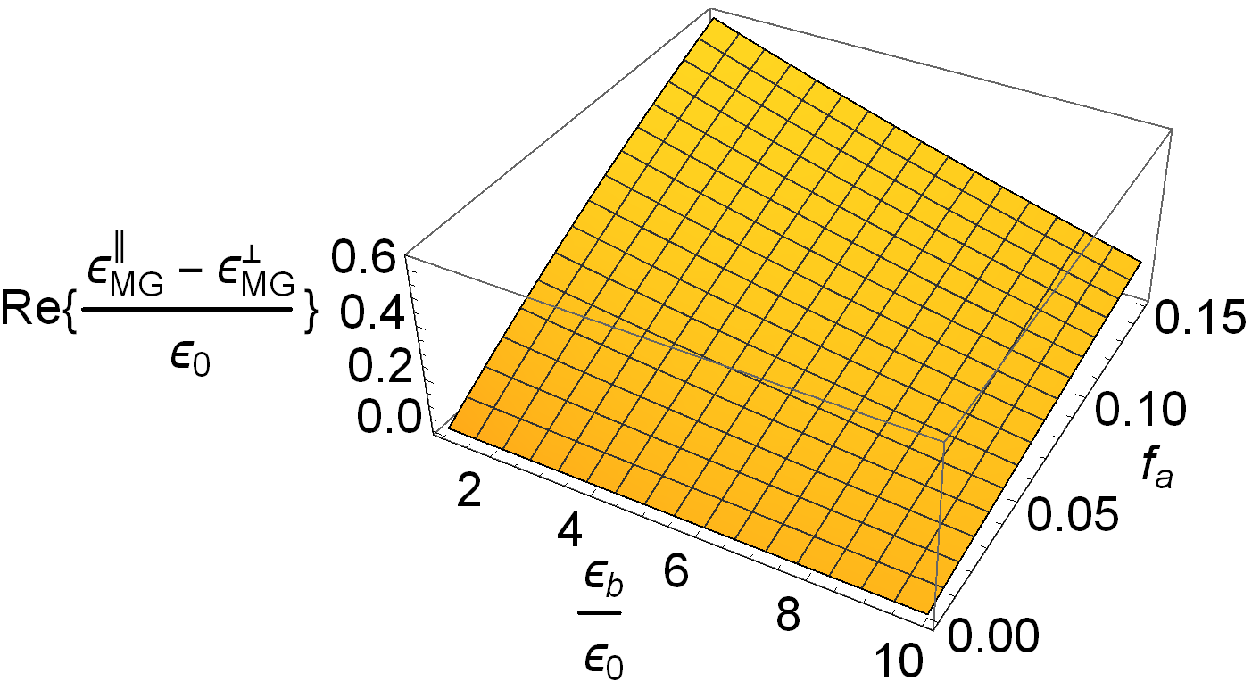,width=3.1in} \hfill
\epsfig{file=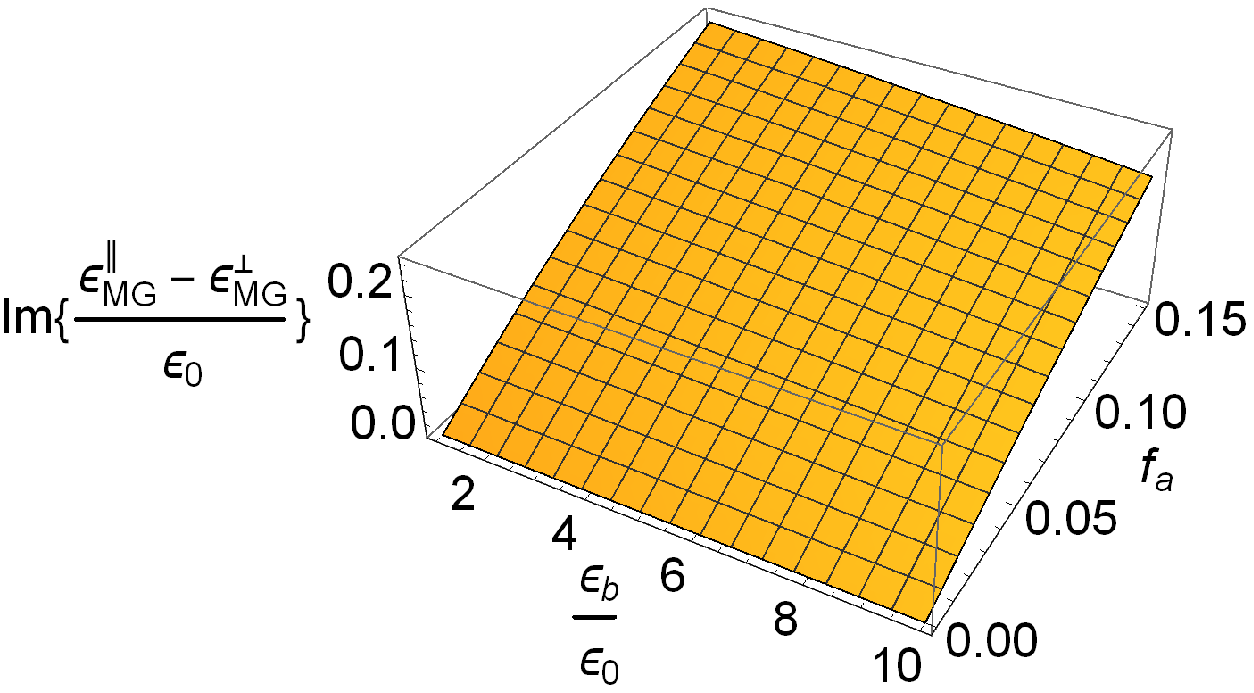,width=3.1in} \vspace{5mm} \\
\epsfig{file=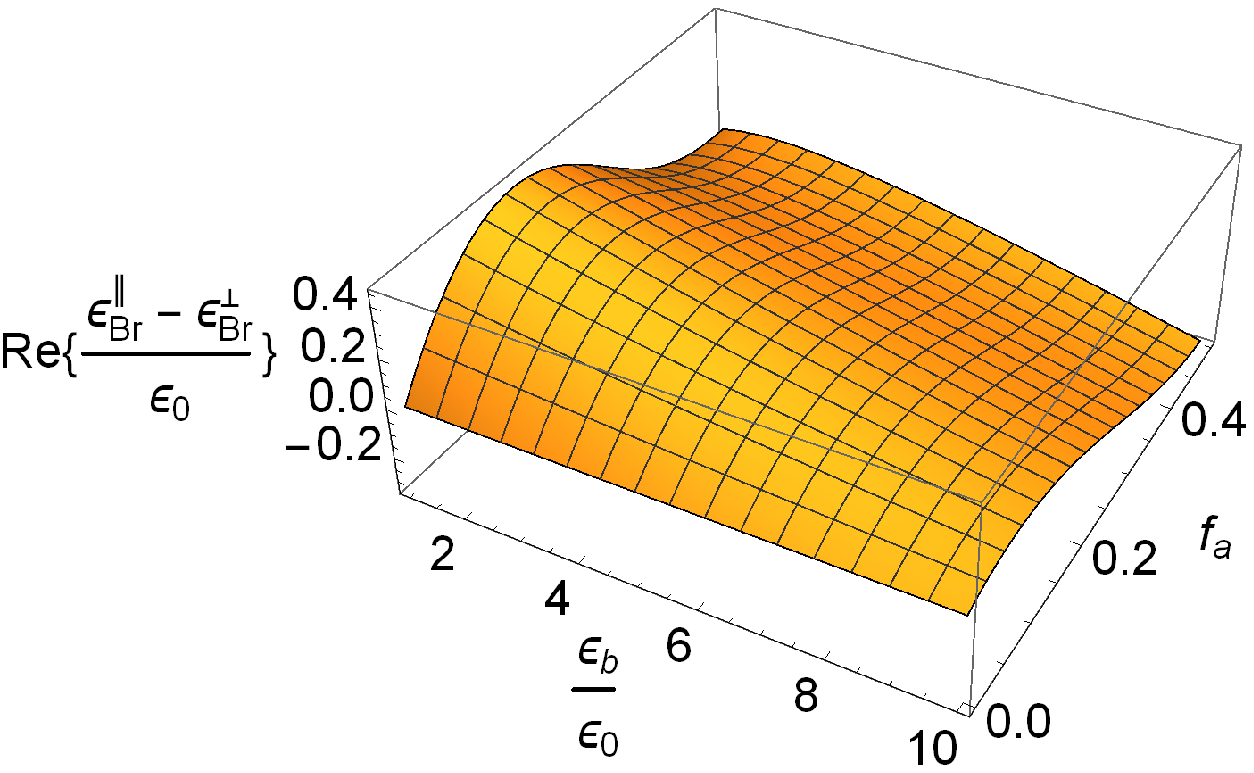,width=3.1in} \hfill
\epsfig{file=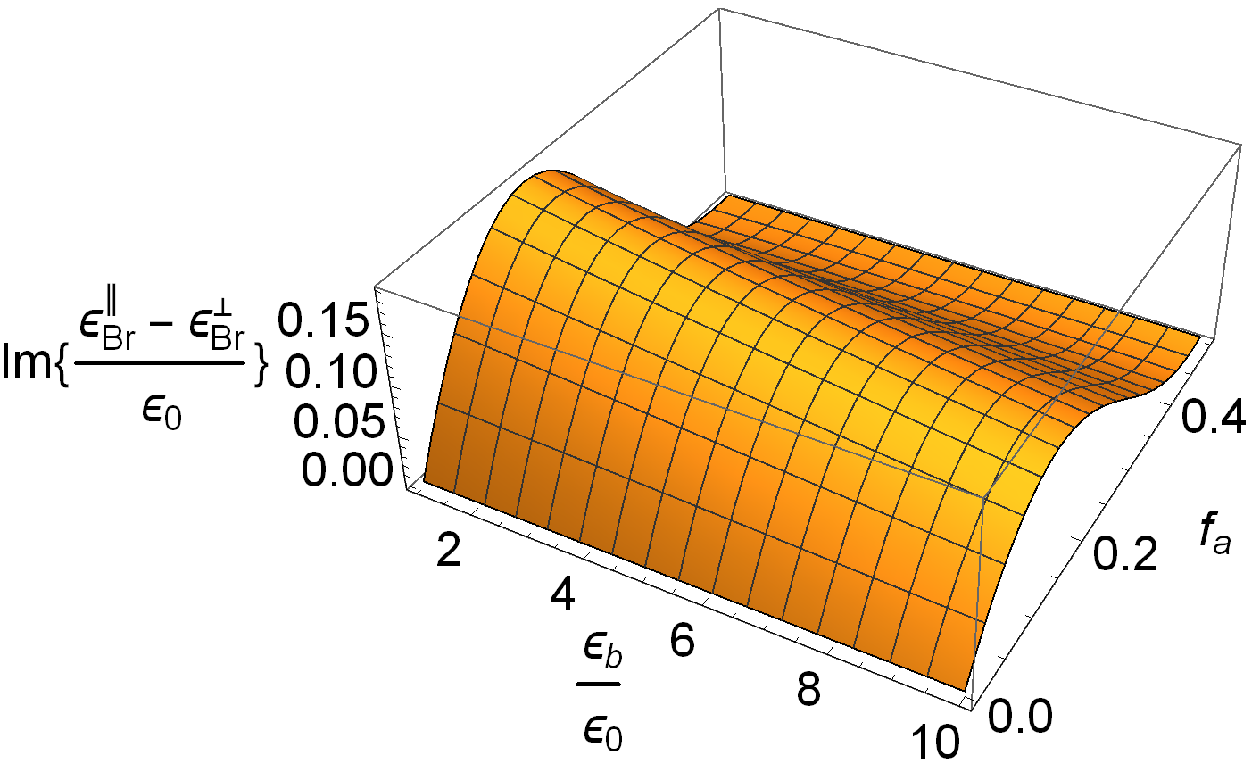,width=3.1in}
\caption{As Fig.~\ref{fig1} except that the dimers are identically oriented and the real and imaginary parts of the differences $\le \eps^{\parallel}_{MG} - \eps^{\perp}_{MG} \ri /\epso$
and $\le \eps^{\parallel}_{Br} - \eps^{\perp}_{Br} \ri/\epso$ are
plotted.
 }
\label{fig3}
\end{figure}

\begin{figure}[!h]
\centering \psfull
 \epsfig{file=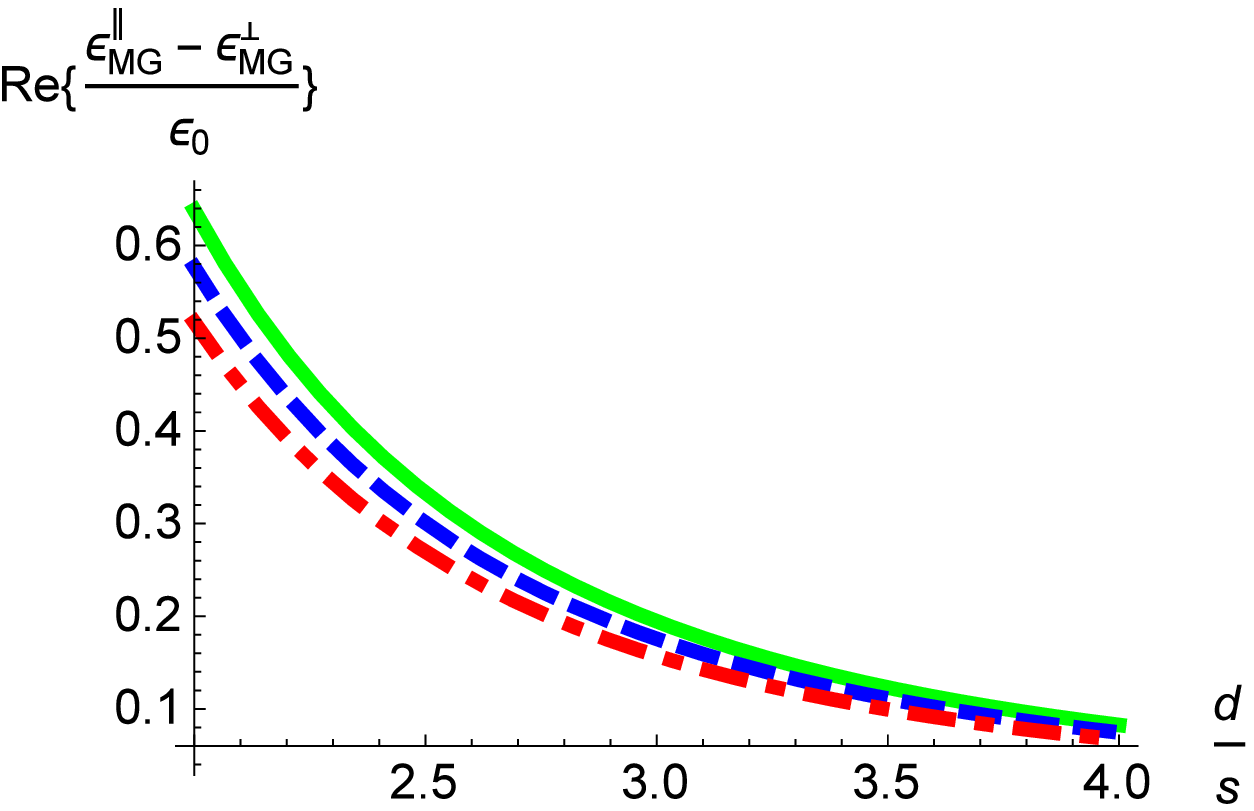,width=3.1in} \hfill
\epsfig{file=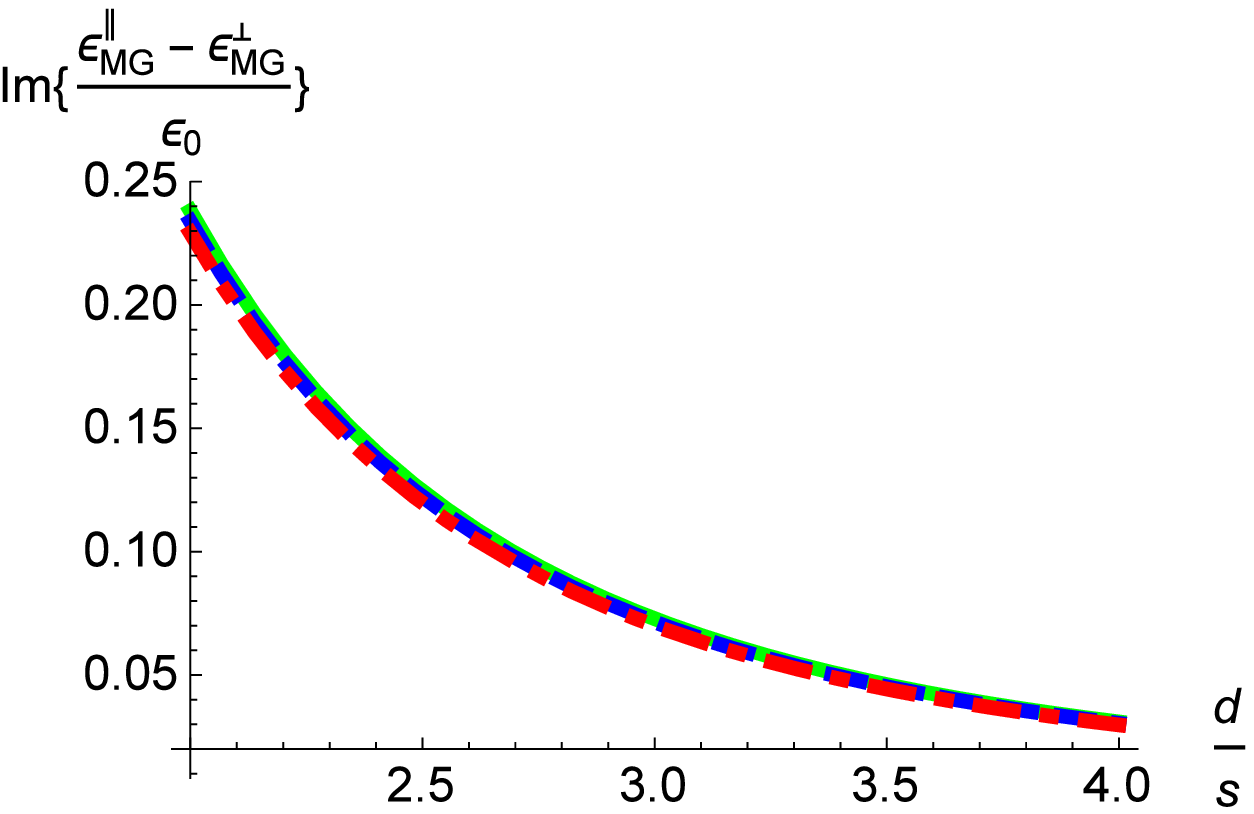,width=3.1in} \vspace{5mm} \\
\epsfig{file=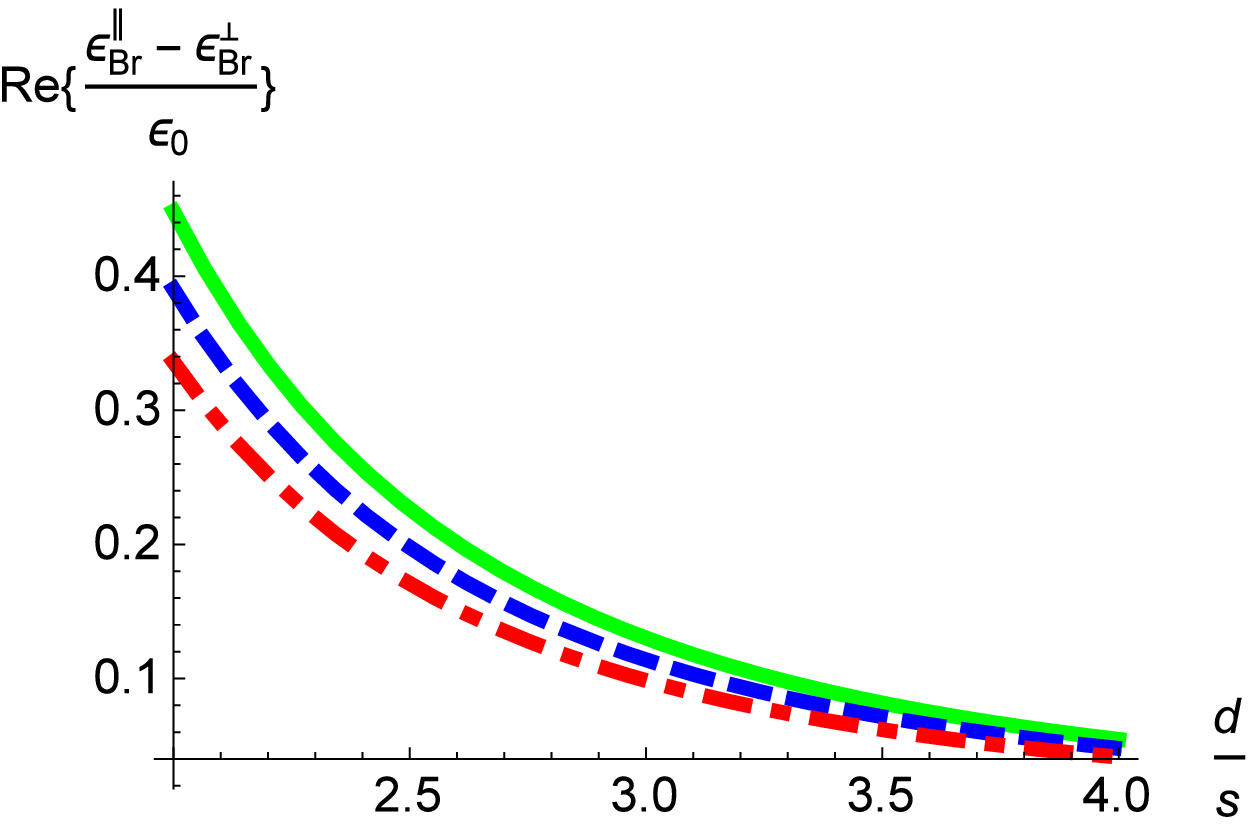,width=3.1in} \hfill
\epsfig{file=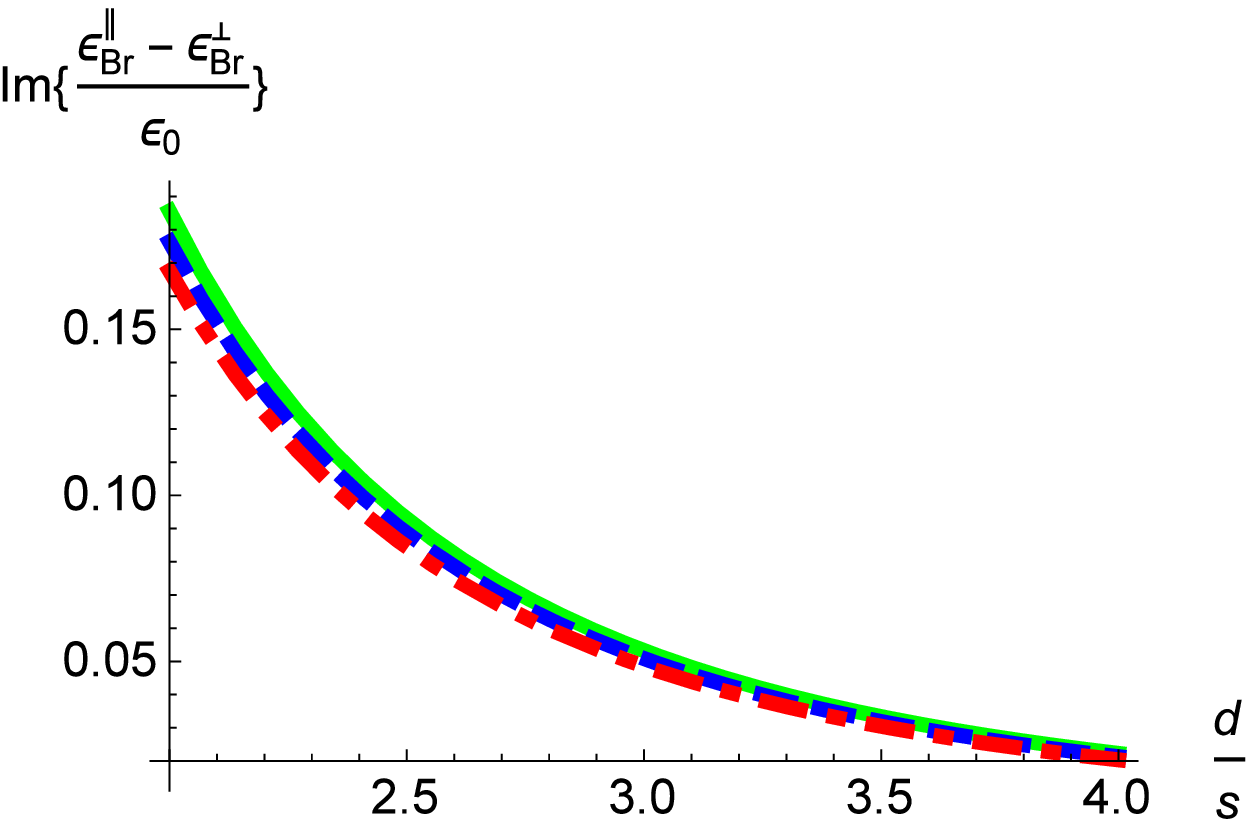,width=3.1in}
\caption{ As Fig.~\ref{fig2} except that  the dimers are identically oriented the real and imaginary parts of
 the differences $\le \eps^{\parallel}_{MG} - \eps^{\perp}_{MG} \ri /\epso$
and $\le \eps^{\parallel}_{Br} - \eps^{\perp}_{Br} \ri/\epso$ are
plotted.
 }
\label{fig4}
\end{figure}

\begin{figure}[!h]
\centering \psfull
 \epsfig{file=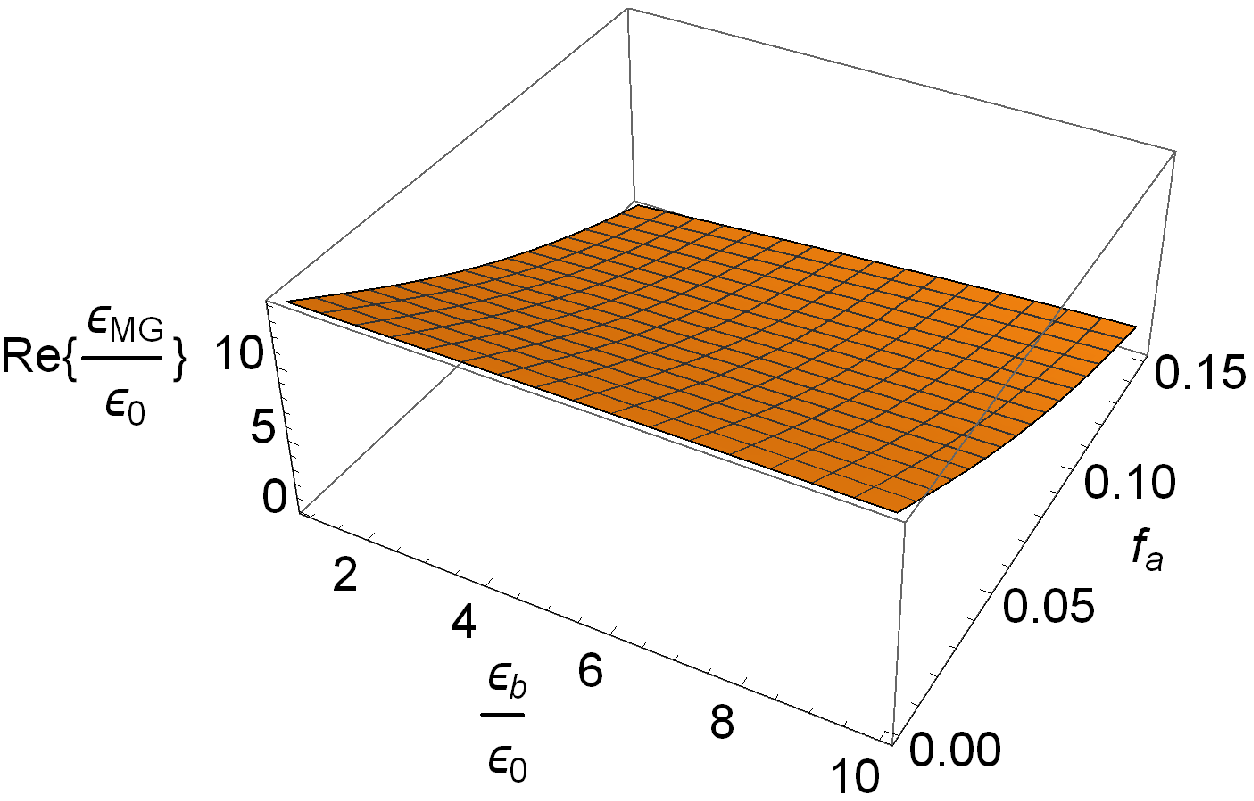,width=3.1in} \hfill
\epsfig{file=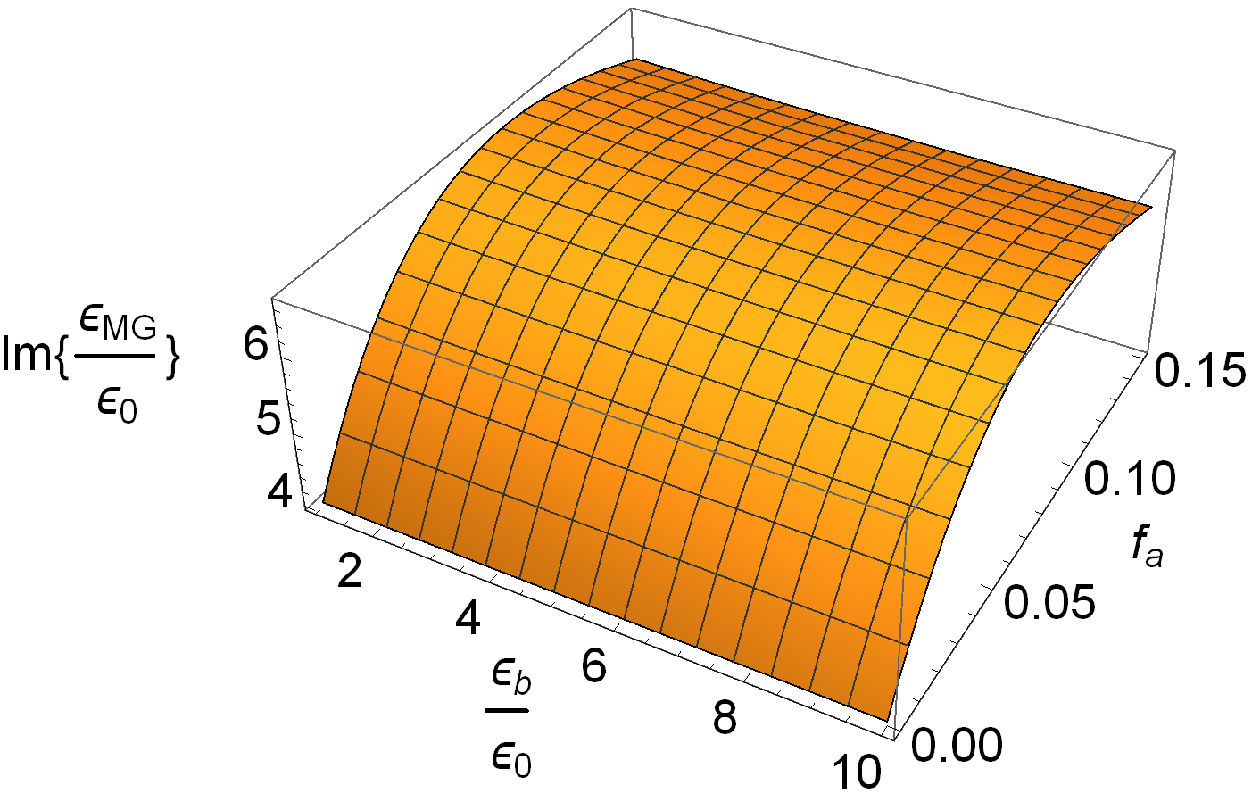,width=3.1in} \vspace{15mm} \\
\epsfig{file=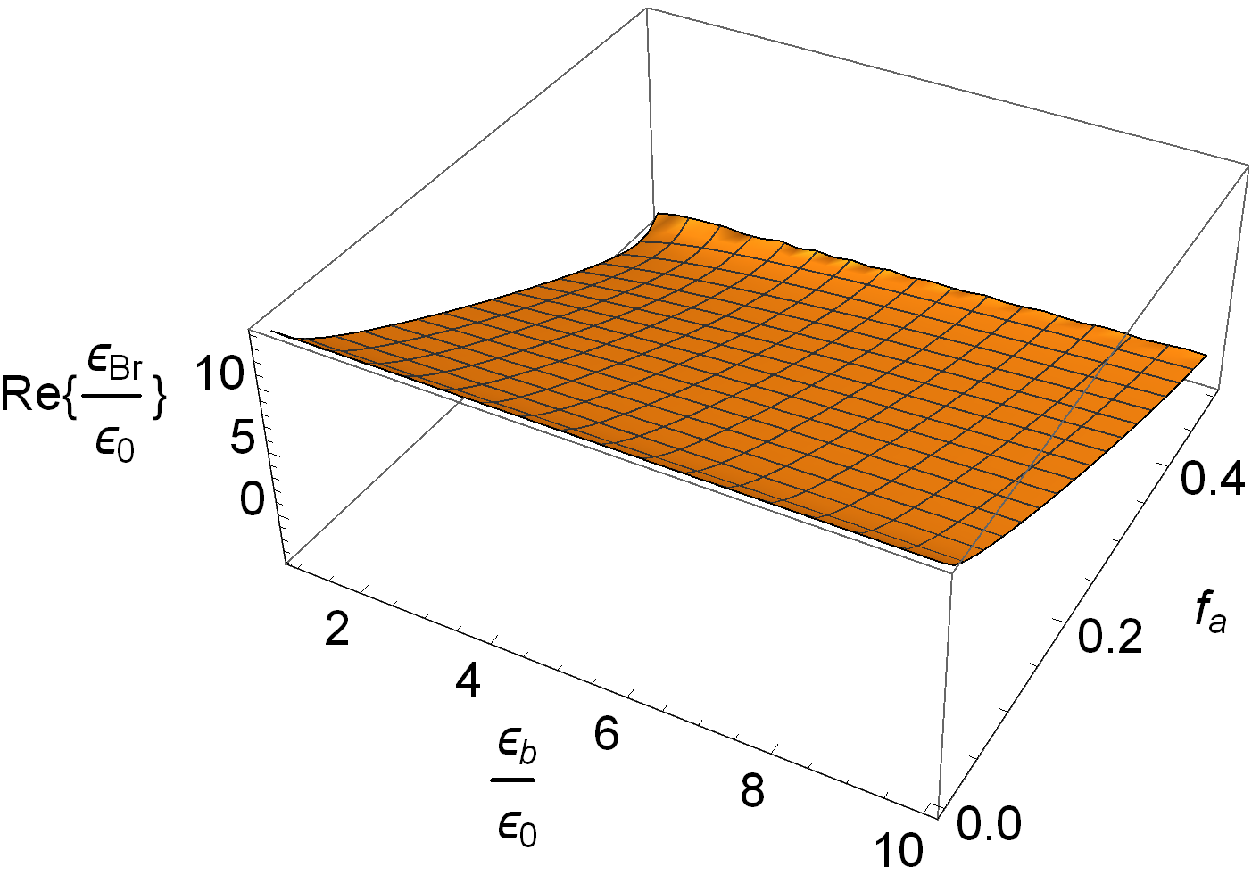,width=3.1in} \hfill
\epsfig{file=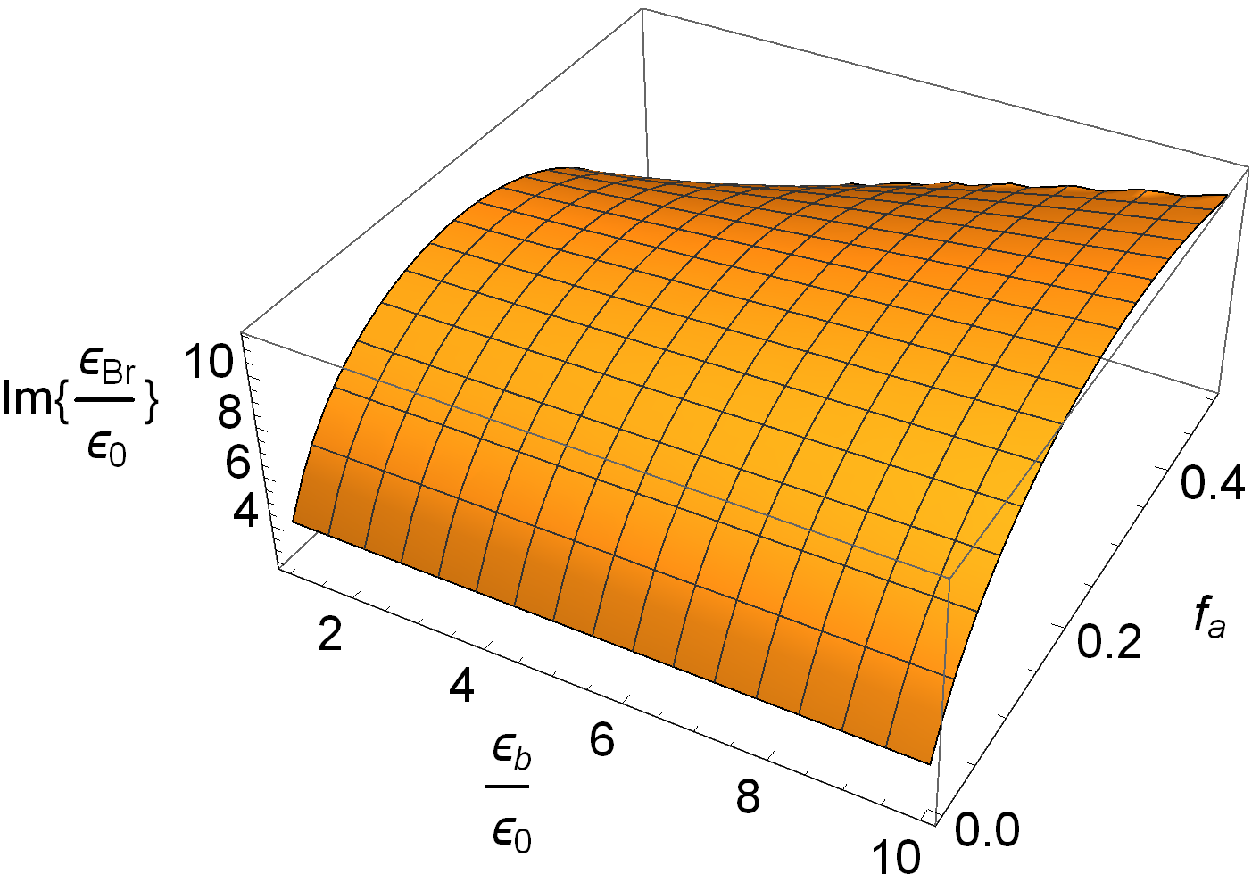,width=3.1in}
\caption{The real and imaginary parts of $\eps_{MG}/\epso$
and $\eps_{Br}/\epso$ plotted against $ \eps_b/\epso $ and $f_a $ for the case where the electrically small spheres of component materials `a' and `b' combine to form metal--dielectric dimers with $\eps_a  = \le -21.4 + 2.4 i \ri \epso$
(i.e., $\eps_a = \eps_{Ag} (5 \, \mbox{nm}) $ for $\lambdao = 650 $ nm) and $\eps_b \in \le 1, 10 \ri \epso$.
 Component material `c' is specified by
$\le \eps_c /\epso \ri = 14 + 4 i$. The dimers are randomly oriented and $d=2s$. }
\label{fig5}
\end{figure}

\begin{figure}[!h]
\centering \psfull
 \epsfig{file=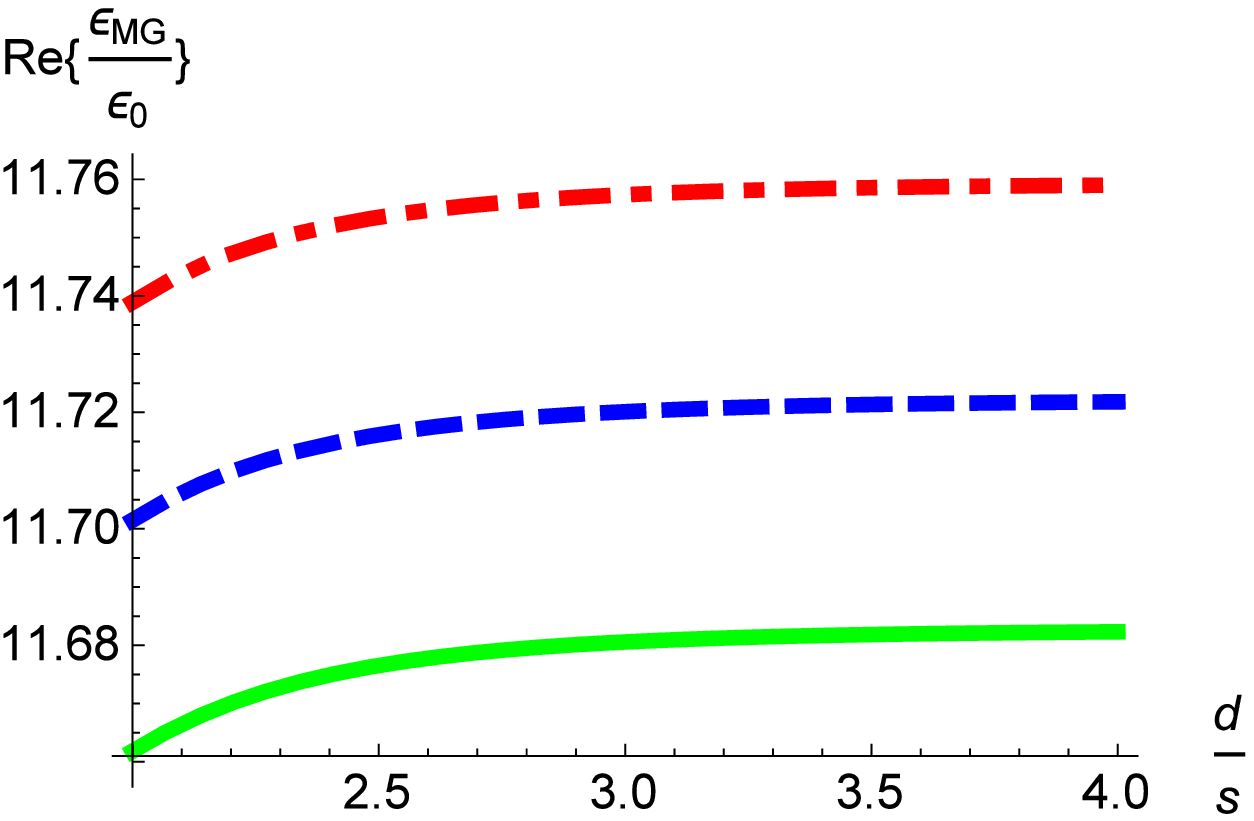,width=3.1in} \hfill
\epsfig{file=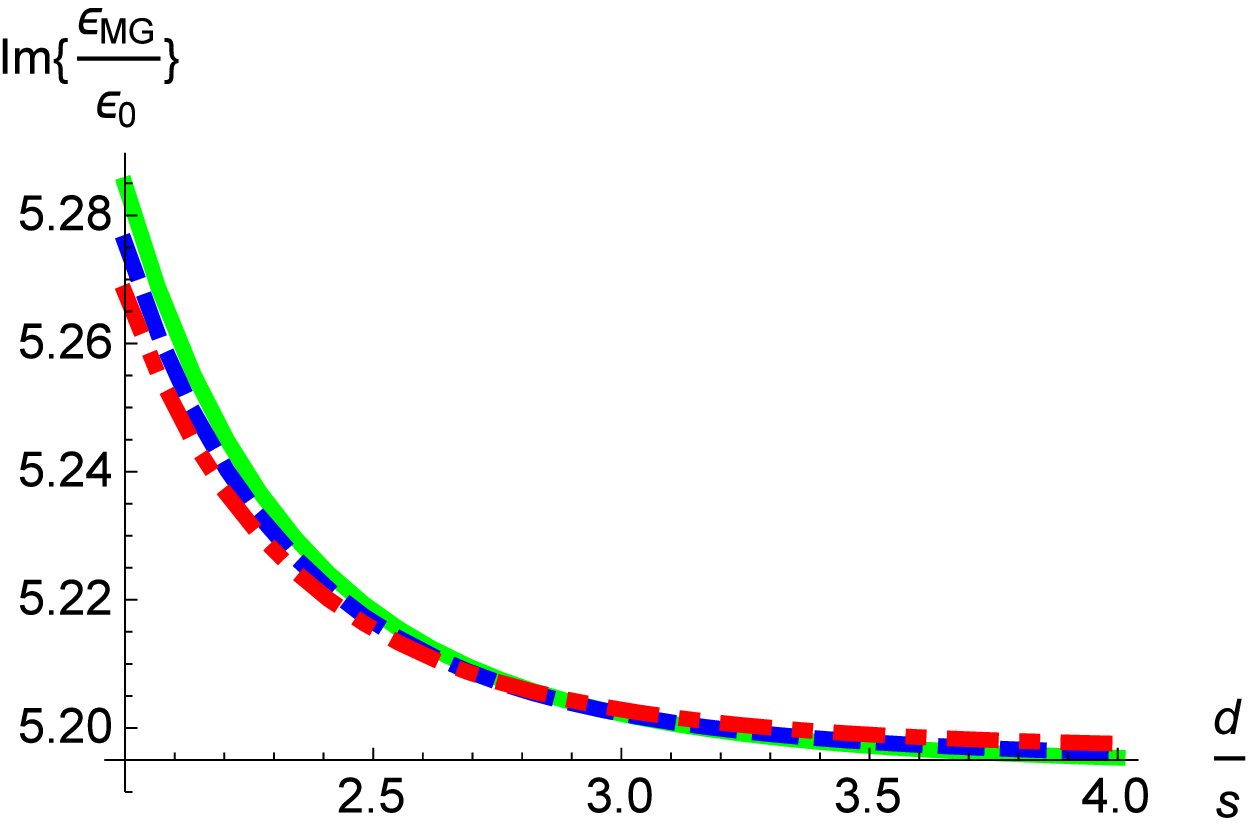,width=3.1in} \vspace{15mm} \\
\epsfig{file=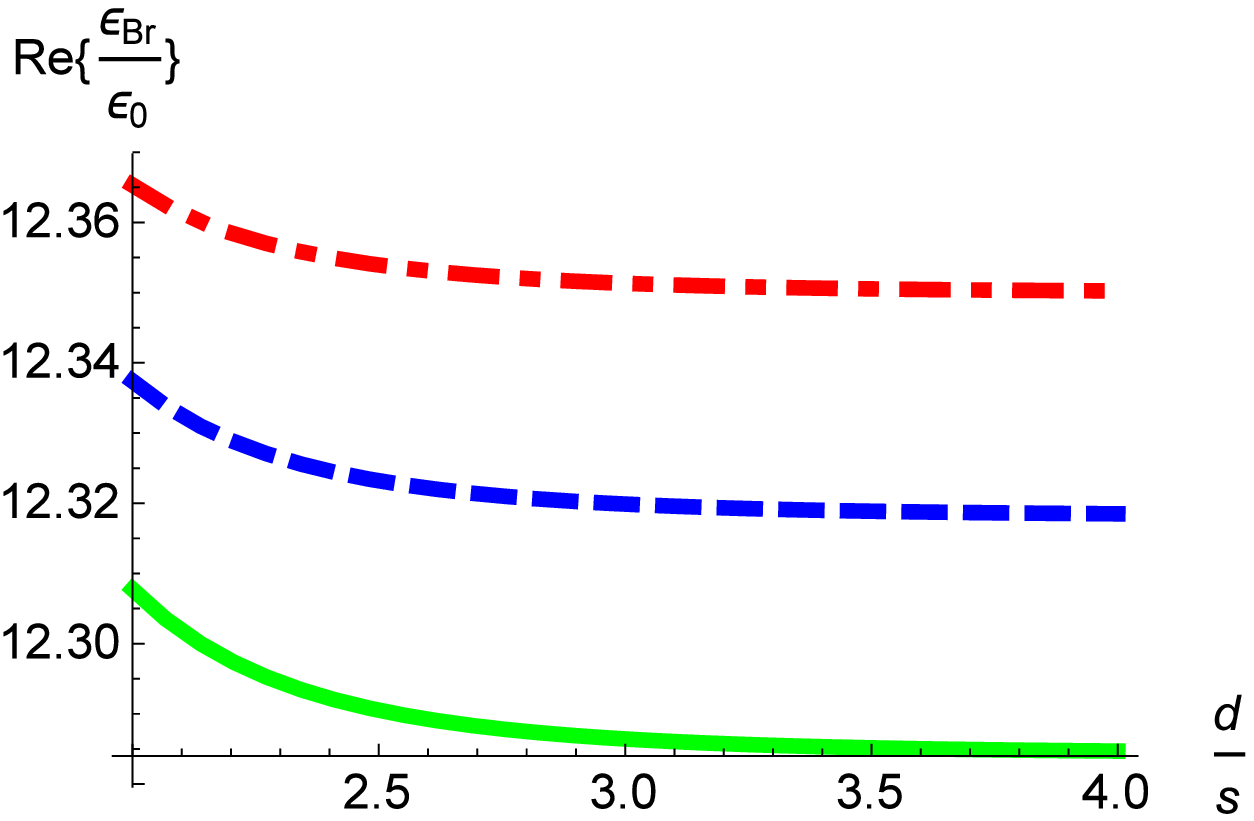,width=3.1in} \hfill
\epsfig{file=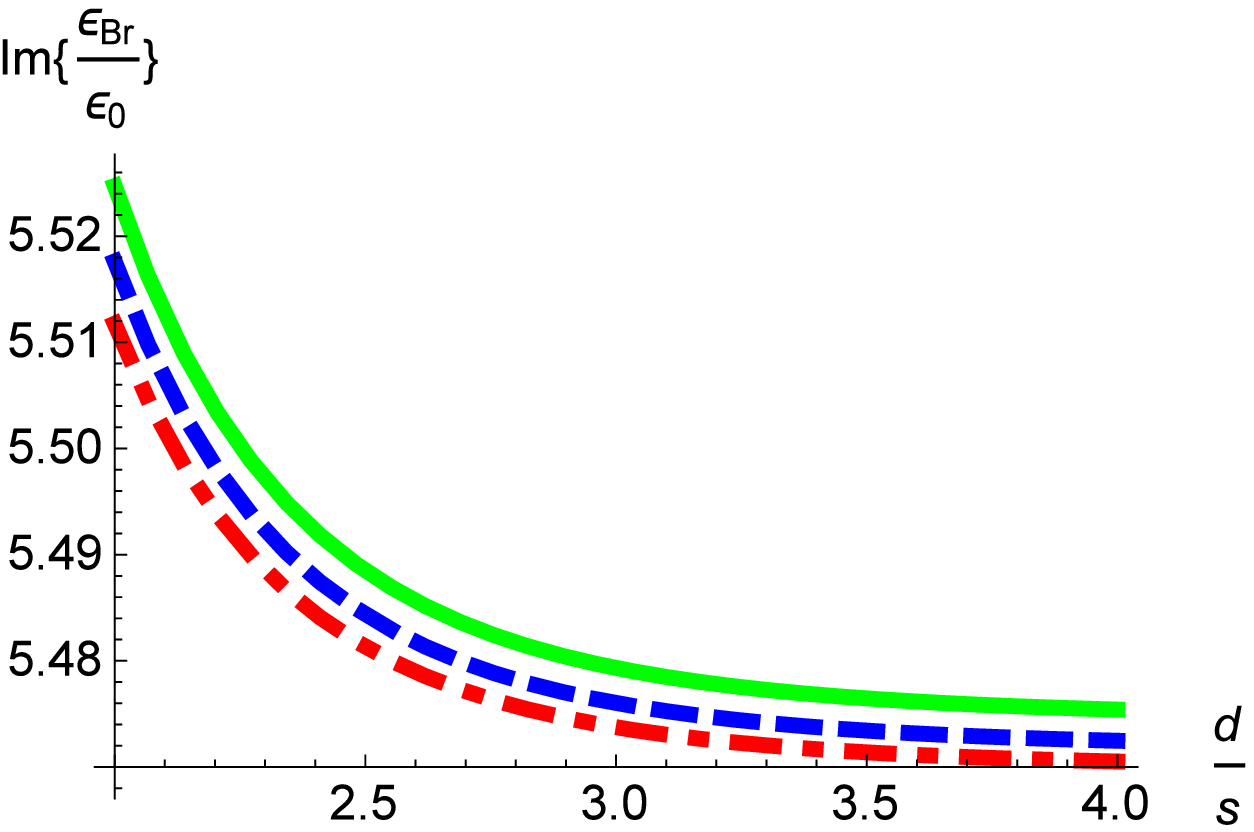,width=3.1in}
\caption{As Fig.~\ref{fig5} except that
the real and imaginary parts of $\eps_{MG}/\epso$
and $\eps_{Br}/\epso$ plotted against  $ d/s $  for  $ \eps_b  = \epso $ (green, solid curves), $2 \epso$ (blue, dashed curves), and $3 \epso$ (red, broken dashed curves). Here $f_a = 0.02$.
 }
\label{fig6}
\end{figure}

\begin{figure}[!h]
\centering \psfull
 \epsfig{file=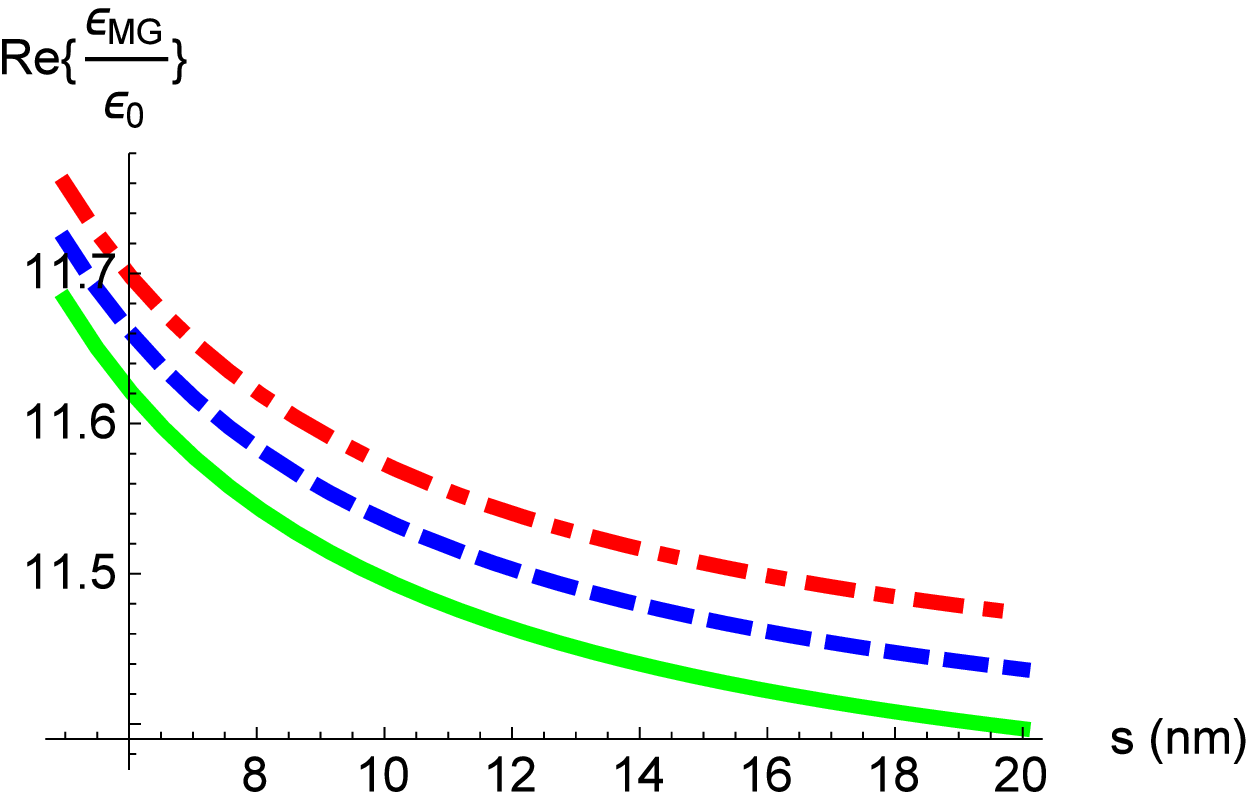,width=3.1in} \hfill
\epsfig{file=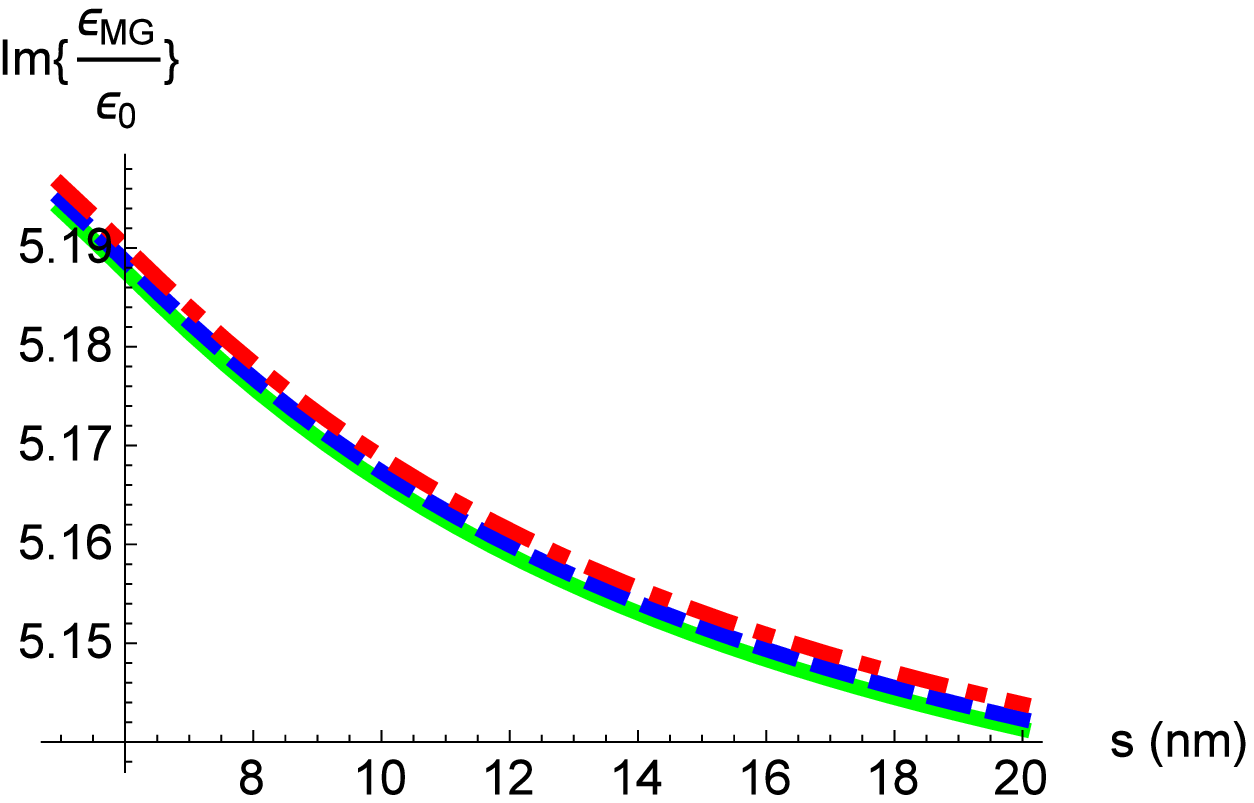,width=3.1in} \vspace{15mm} \\
\epsfig{file=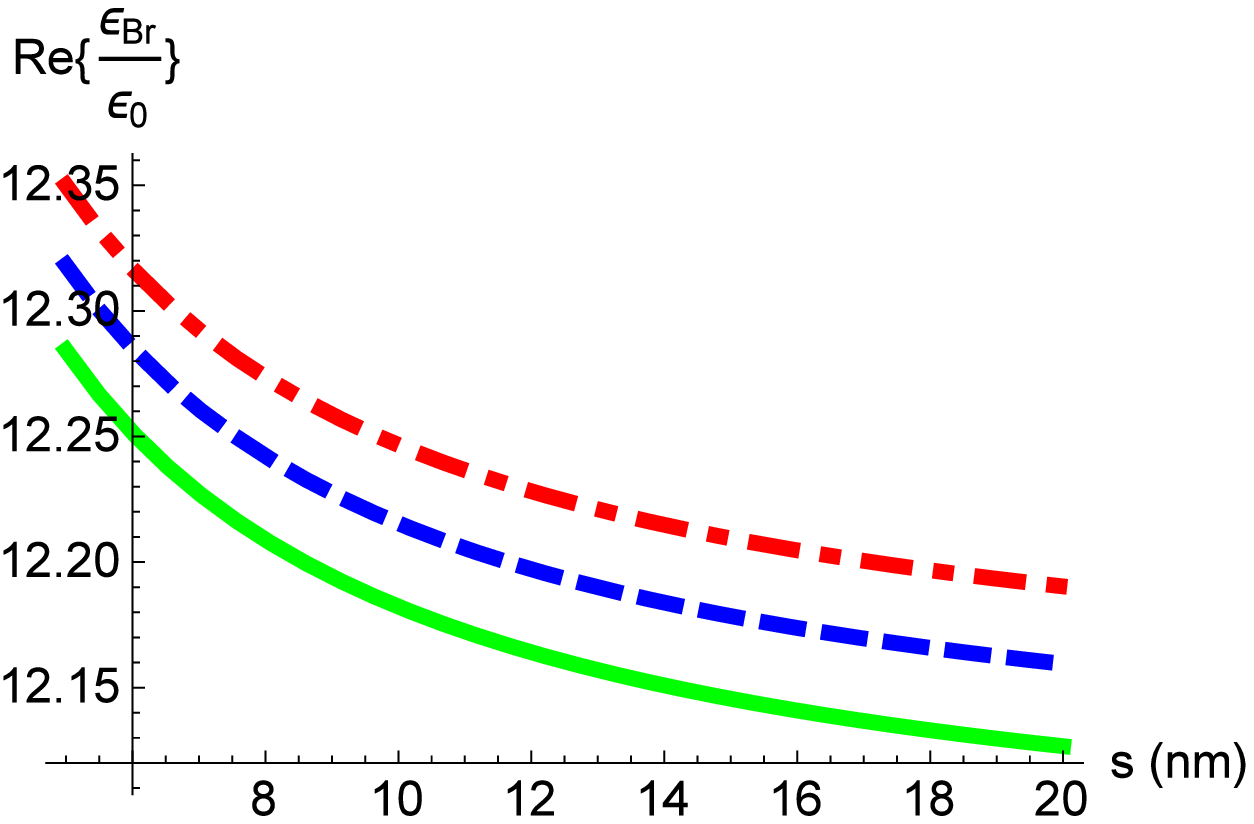,width=3.1in} \hfill
\epsfig{file=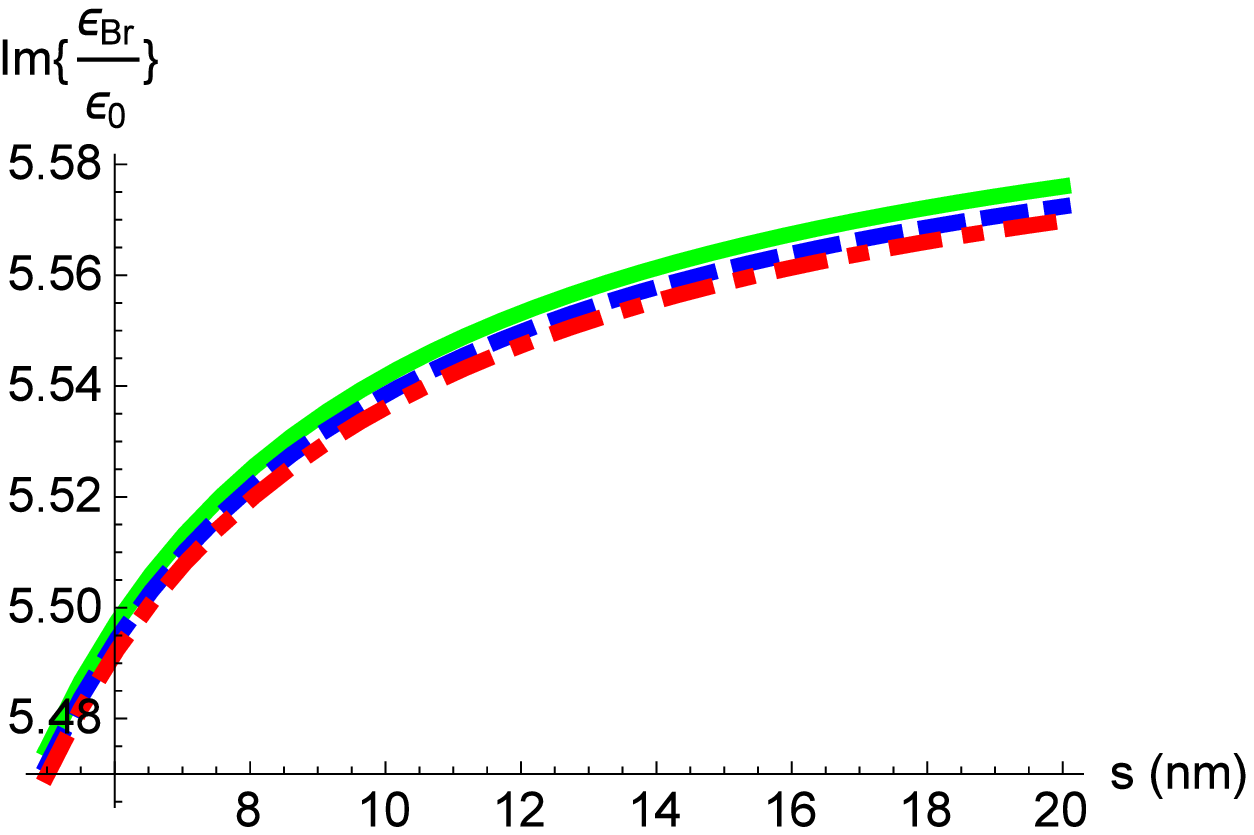,width=3.1in}
\caption{As Fig.~\ref{fig6} except that
the real and imaginary parts of $\eps_{MG}/\epso$
and $\eps_{Br}/\epso$ are plotted against  $  s $ (nm) with $d = 2 s$.
 }
\label{fig7}
\end{figure}

\begin{figure}[!h]
\centering \psfull
 \epsfig{file=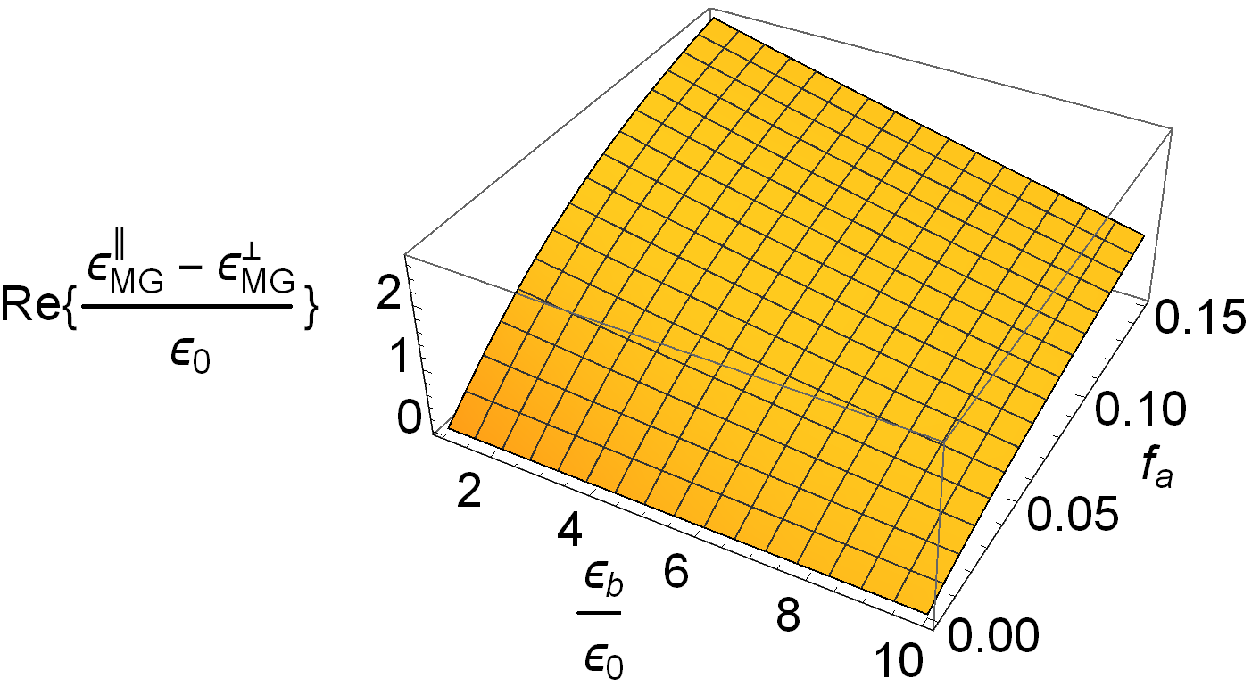,width=3.1in} \hfill
\epsfig{file=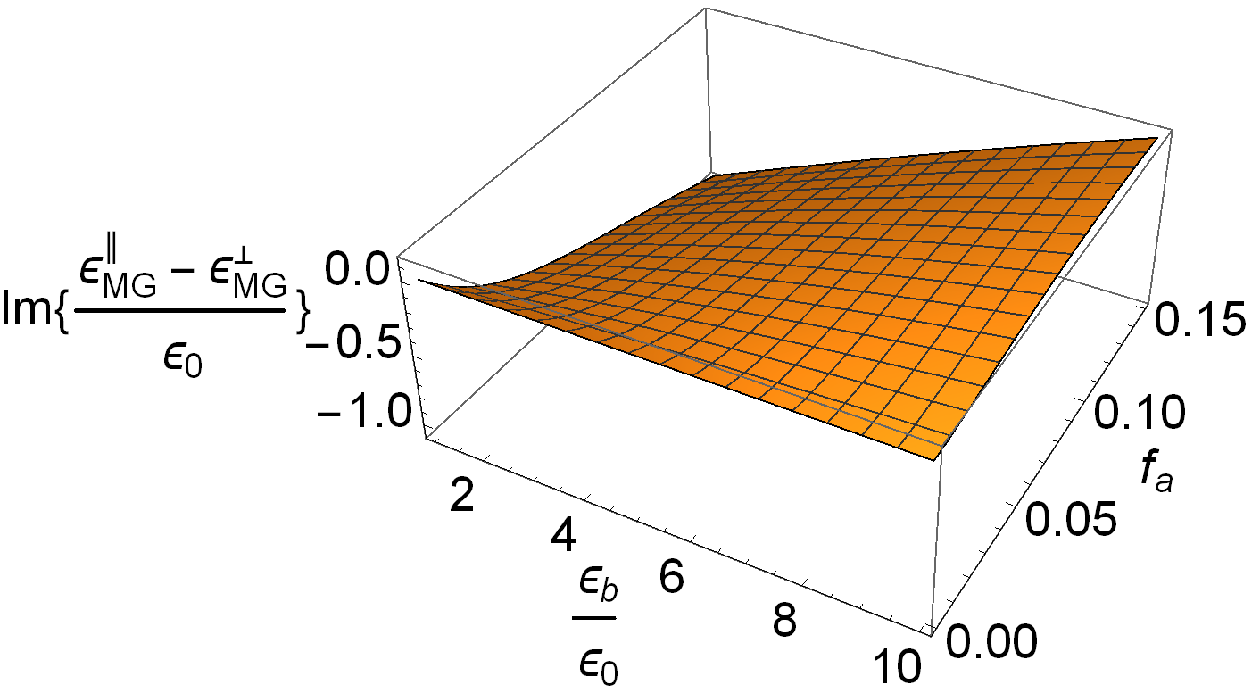,width=3.1in} \vspace{5mm} \\
\epsfig{file=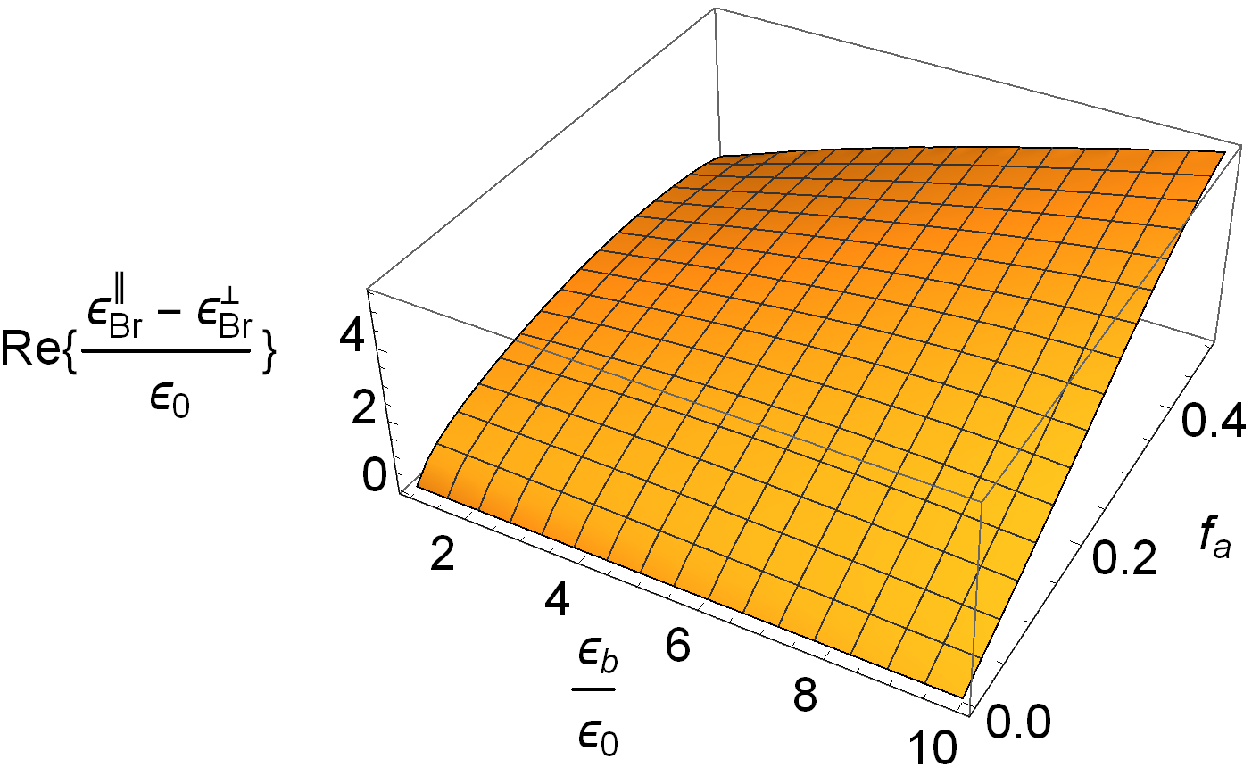,width=3.1in} \hfill
\epsfig{file=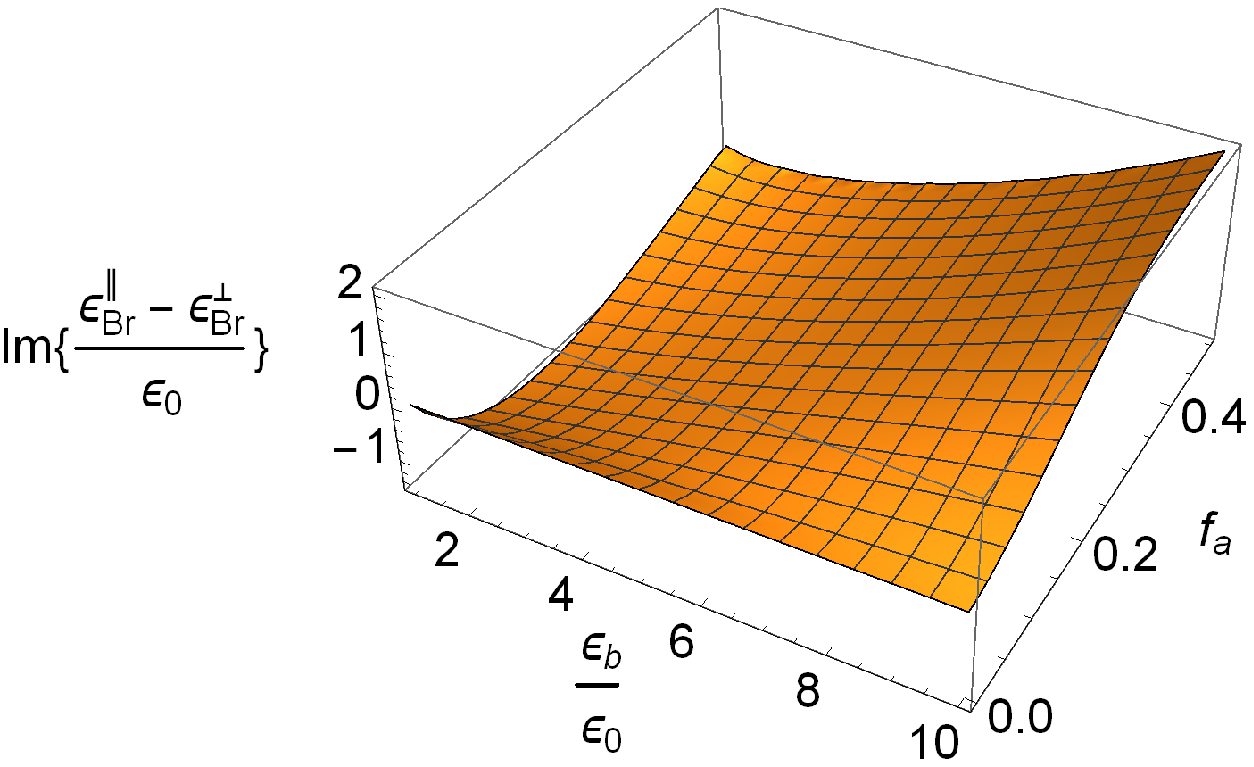,width=3.1in}
\caption{As Fig.~\ref{fig5} except that the dimers are identically oriented and the real and imaginary parts of the differences $\le \eps^{\parallel}_{MG} - \eps^{\perp}_{MG} \ri /\epso$
and $\le \eps^{\parallel}_{Br} - \eps^{\perp}_{Br} \ri/\epso$ are
plotted.
 }
\label{fig8}
\end{figure}

\begin{figure}[!h]
\centering \psfull
 \epsfig{file=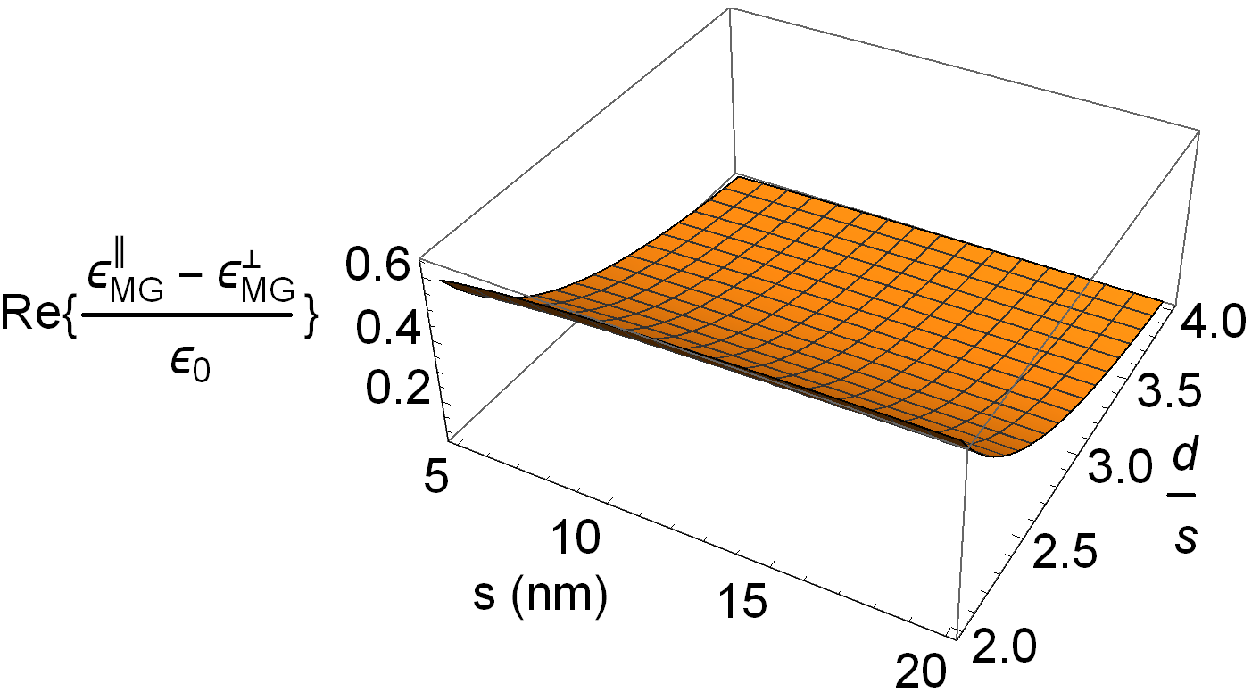,width=3.1in} \hfill
\epsfig{file=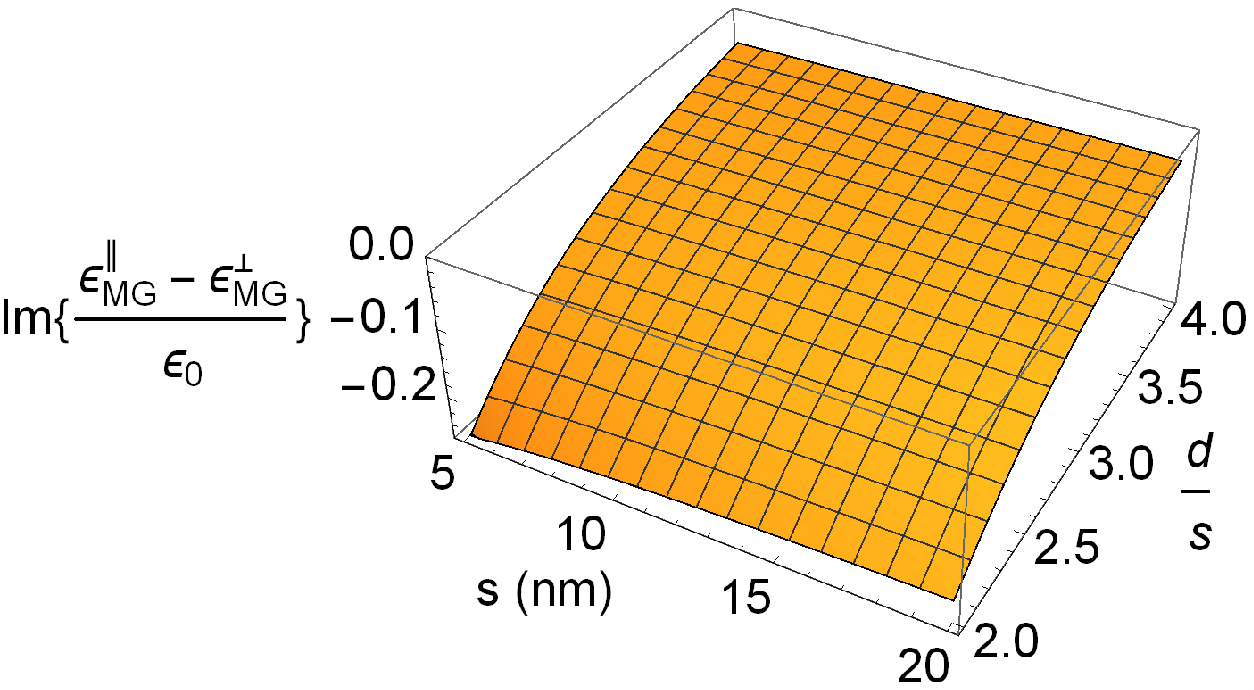,width=3.1in} \vspace{5mm} \\
\epsfig{file=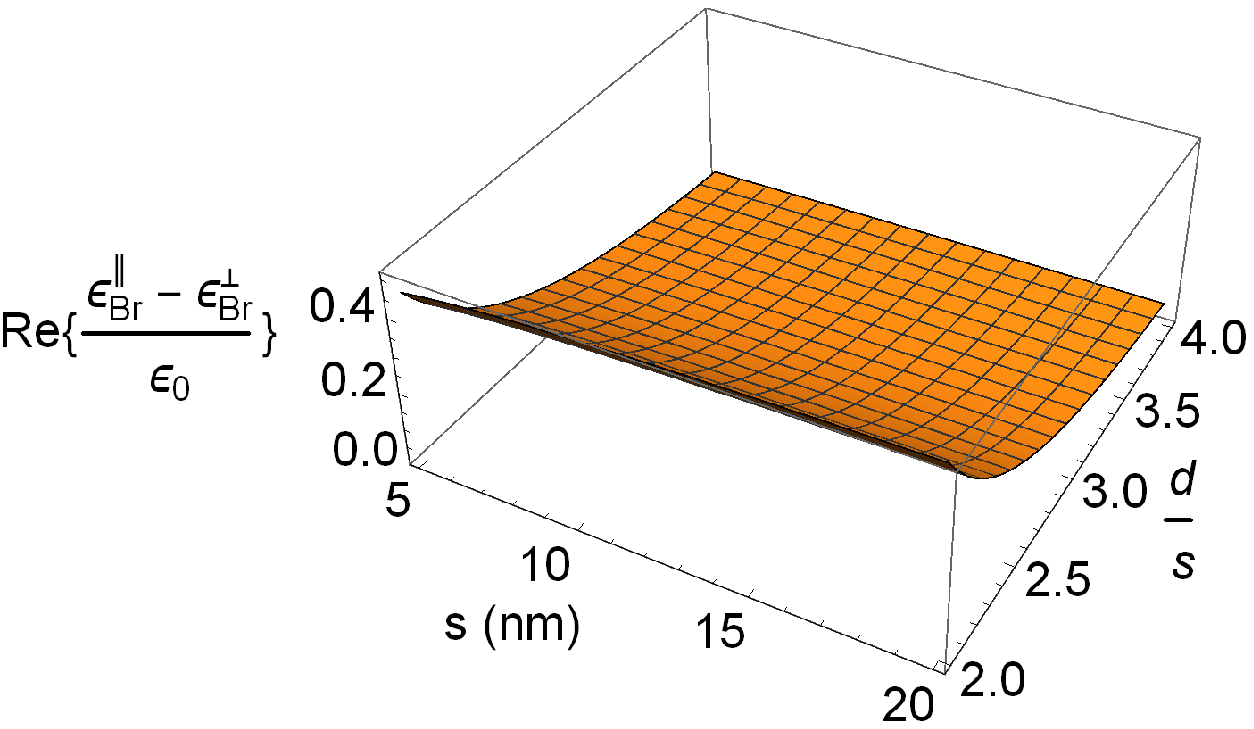,width=3.1in} \hfill
\epsfig{file=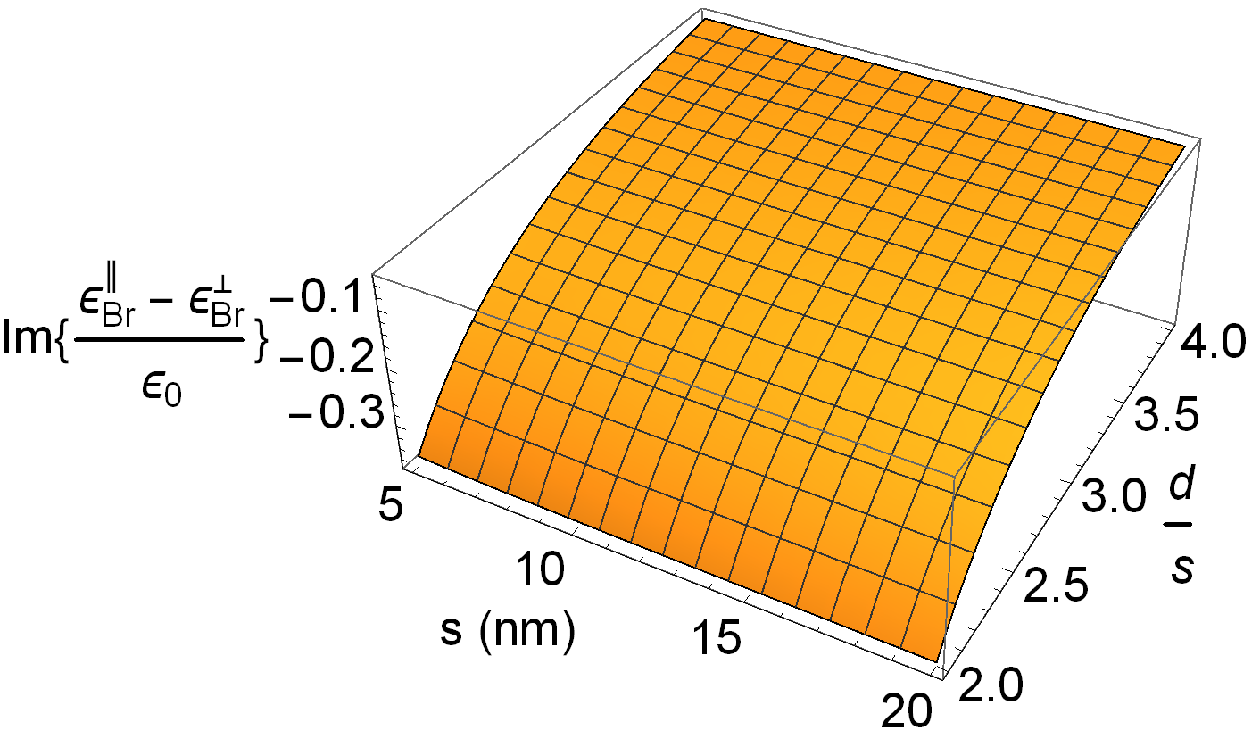,width=3.1in}
\caption{As Fig.~\ref{fig8} except that $\le \eps^{\parallel}_{MG} - \eps^{\perp}_{MG} \ri /\epso$
and $\le \eps^{\parallel}_{Br} - \eps^{\perp}_{Br} \ri/\epso$ are
plotted against $s$ (nm) and $d/s$. Here $\eps_b = 2 \epso$ and  $f_a = 0.02$.
 }
\label{fig9}
\end{figure}

The numerical results in \S\ref{Num_sec} reveal that the effects of  intradimer coupling decay rapidly as $d$ increases. Indeed, for $d > 4s$ these effects are generally negligible and the permittivity dyadic of the HCM is practically the same as that which would arise in the case
where component materials `a' and `b' were not coupled at all. By comparing the numerical results at $d = 2s$ with those at $d = 4 s$ in Figs.~\ref{fig2}, \ref{fig4}, \ref{fig6}, and \ref{fig9}, it may be deduced that
intradimer coupling generally has  relatively modest but not insignificant effects on the HCM parameter estimates delivered by the Bruggeman and Maxwell Garnett formalisms, and that these effects are exacerbated by anisotropy in the case of identically oriented dimers.

\vspace{5mm}
\noindent {\bf Acknowledgement.  } 
AL thanks the Charles Godfrey Binder Endowment at Penn State for ongoing support of his research activities.

\newpage

\newpage

\newpage

\newpage

\newpage

\newpage

\newpage

\newpage

\newpage

\newpage

\begin{thebibliography}{99}


\bibitem{Nanoengineering}
Schulz MJ,  Kelkar AD, Sundaresan MJ.
Nanoengineering of structural, functional and smart materials. London (UK): CRC Press; 2005.

\bibitem{Lombardi}
 Lombardi A,  Grzelczak MP,  Crut A,  Maioli P,  Pastoriza-Santos I,
 Liz-Marz\'an LM,  Del Fatti N,  Vall\'ee F.
Optical response of individual
Au-Ag@SiO$_2$ heterodimers.
ACS Nano 2013;  7:2522--2531.

\bibitem{Park}
Park W.
Optical interactions in plasmonic nanostructures.
Nano Convergence 2014; 1:2.

\bibitem{Nordlander}
 Marinica DC, Kazansky AK,  Nordlander P, Aizpurua J, Borisov AG.
Quantum plasmonics: nonlinear effects in the field enhancement
of a plasmonic nanoparticle dimer.
Nano Lett. 2012: 12:1333--1339.

\bibitem{Mott_insulator_PRL}
Okazaki R,  Ikemoto Y,  Moriwaki T,  Shikama T,  Takahashi K,  Mori H,  Nakaya H,  Sasaki T,  Yasui Y,  Terasaki I.
Optical conductivity measurement of a dimer Mott-insulator to charge-order phase transition in a two-dimensional
quarter-filled organic salt compound.
Phys. Rev. Lett. 2013;  111:217801.

\bibitem{Mott_insulator_PRB}
 Chen R,  Lee SB,  Balents L.
Dimer Mott insulator in an oxide heterostructure.
Phys. Rev. B  2013; 87:161119.

\bibitem{Dimer_Liquid_Crystals}
Gupta SK, Singh DP,  Manohar R,  Hiremath US,
 Yelmaggad CV.
Dielectric behaviour of a ferroelectric liquid crystal dimer.
Liquid Crystals 2012; 39:1125--1129.

\bibitem{Hydrogen_dimer}
 \v{S}ljivan\v{c}anin \v{Z},  Rauls E,  Hornek{\ae}r L,  Xu W,  Besenbacher F,  Hammer B.
Extended atomic hydrogen dimer configurations on the graphite(0001) surface.
J. Chem. Phys. 2009; 131:084706.

\bibitem{Zhang}
Zhang F,  Sadaune V,  Kang L,  Zhao Q,  Zhou J,
 Lippens D. Coupling effect for dielectric metamaterial dimer.
Prog. Electromag. Res. 2012; 132:587--601.

\bibitem{Torquato}
Kim IC, Torquato S. Effective conductivity of suspensions of overlapping spheres.
J. Appl. Phys. 1992; 71:2727--2735.

\bibitem{L96}  Lakhtakia A (ed.).
Selected papers on linear optical composite materials. Bellingham (WA): SPIE
Optical Engineering Press; 1996.

\bibitem{ML_PiO}
 Mackay TG,  Lakhtakia A. Electromagnetic fields in linear bianisotropic mediums. Prog. Opt. 2008; 51:121--209.

\bibitem{TK81}  Tsang L, Kong JA.
Scattering of electromagnetic waves from random media with strong permittivity fluctuations.
Radio Sci. 1981;  16:303--320. 

\bibitem{Faxen}
Fax\'{e}n H.
Der Zusammenhang zwischen den Maxwellschen Gleichungen
f\"{u}r Dielektrika und den atomistischen Ans\"{a}tzen von H. A.
Lorentz u.a.
Zeit. Phys. 1920; 2:218--229.
	
\bibitem{Lakh1992}
Lakhtakia A.
Size-dependent Maxwell-Garnett formula from an integral equation formalism. Optik
1992; 91:134--137.

\bibitem{H-S_bounds}  Hashin Z,   Shtrikman S.
 A variational approach to the theory of the effective magnetic
permeability of multiphase materials. J. Appl. Phys. 1962; 33:3125--3131. 

\bibitem{Chen}
 Chen HC. Theory of electromagnetic waves. New York (NY):  McGraw--Hill;
 1983.


\bibitem{EAB}
Mackay TG, Lakhtakia A.  Electromagnetic anisotropy and bianisotropy.
Singapore: World Scientific; 2010.

\bibitem{Jackson}
 Jackson JD. Classical electrodynamics, 3rd edn.  New
York (NY): Wiley;   1999.


\bibitem{Michel}  Michel B. A Fourier space approach to the pointwise
singularity of an anisotropic dielectric medium. Int. J. Appl.
Electromagn. Mech. 1997;   8:219--227.

\bibitem{WLM1993}  Weiglhofer WS,  Lakhtakia A,   Monzon JC.
Maxwell--Garnett model for composites of electrically small uniaxial objects.
Microw. Opt. Technol. Lett. 1993; 6:681--684.



\bibitem{RL2005}
Ross BM,  Lakhtakia A.
Bruggeman approach for isotropic chiral mixtures revisited.
  Microw. Opt. Technol. Lett. 2005; 44:524--527.

\bibitem{WLM97}  Weiglhofer WS,  Lakhtakia A,   Michel B.
  Maxwell Garnett and Bruggeman formalisms for a particulate composite with bianisotropic host medium.
  Microw. Opt. Technol. Lett. 1997; 15:263--266.
Corrections: 1999;  22:221. 

\bibitem{Jacobi}
Rao SS.
Applied numerical methods for engineers and scientists. Cambridge (UK): Pearson Publishing; 2001.


\bibitem{BH}
  Bohren CF,    Huffman DR.  Absorption and scattering of light
by small particles. New York (NY): Wiley; 1983.

\bibitem{Kreibig}
 Kreibig U. Electronic properties of small silver particles: the optical constants and their temperature dependence.
J. Phys. F: Metal Phys. 1974;  4:999--1014.


\bibitem{ML_Bruggeman}
 Mackay TG,  Lakhtakia A.
 A limitation of the Bruggeman formalism for homogenization. Opt. Commun. 2004;  234:35--42.
 Corrections: 2009; 282:4028. 

\bibitem{M_JNP}
 Mackay TG. On the effective permittivity of silver--insulator nanocomposites.
 J. Nanophoton. 2007; 1:019501. 

\bibitem{Bro1}
Fourn C,  Brosseau C.
Electrostatic resonances of heterostructures with negative permittivity:
Homogenization formalisms versus finite-element modeling.
Phys. Rev. E 2008; 77:016603.

\bibitem{Bro2}
Mejdoubi A,  Brosseau C.
Electrostatic resonance of clusters of dielectric cylinders:
A finite element simulation.
Phys. Lett. A 2008; 372:741--748.

\bibitem{Limit_active}
 Mackay TG,   Lakhtakia A.
 On the application of  homogenization formalisms to active dielectric composite materials.
 Opt. Commun. 2009; 282:2470--2475. 


\bibitem{Limit_inverse}
 Jamaian SS,   Mackay TG.
 On limitations of the  Bruggeman formalism for inverse homogenization.
J. Nanophoton. 2010;  4:043510. 

\bibitem{Limit_BM}
 Duncan AJ,    Mackay TG,   Lakhtakia A.
 On the Bergman--Milton bounds for the homogenization of dielectric composite materials.
 Opt. Commun. 2007; 271:470--474. 


\end{thebibliography}
\end{document}